\documentclass{article}

\PassOptionsToPackage{numbers, compress}{natbib}
\usepackage[final]{tackling_climate_workshop_style}

\usepackage[utf8]{inputenc} \usepackage[T1]{fontenc}    \usepackage{float}       \usepackage{hyperref}       \usepackage{url}            \usepackage{booktabs}       \usepackage{amsfonts}       \usepackage{nicefrac}       \usepackage{microtype}      \usepackage{xcolor}
\usepackage{graphicx}
\usepackage{xfrac}

\usepackage{doi}

\usepackage{color, colortbl}
\definecolor{darkblue}{rgb}{0.0,0.0,0.65}
\definecolor{darkred}{rgb}{0.68,0.05,0.0}
\definecolor{darkgreen}{rgb}{0.0,0.29,0.29}
\definecolor{darkpurple}{rgb}{0.47,0.09,0.29}
\definecolor{myblue}{rgb}{0.1380392156862745, 0.3027450980392157, 0.6274509803921569}
\definecolor{myred}{rgb}{0.6235294117647059, 0.13725490196078433, 0.0196078431372549} 
\hypersetup{
  colorlinks = true,
  citecolor  = myblue,
  linkcolor  = myred,
  filecolor  = darkblue,
  urlcolor   = darkblue,
 }

\usepackage{etoolbox}
\usepackage[normalem]{ulem}
\definecolor{revgreen}{rgb}{0.0,0.55,0.0}
\newtoggle{showrev}
\togglefalse{showrev}

\graphicspath{{figures/}}

\title{Variable Masking for Climate Emulation in ACE}

\author{Alexey Yermakov$^{1,2}$\thanks{2026 summer intern at Ai2, \texttt{alexeyy@uw.edu}} \space, Troy Arcomano$^{1}$, Jeremy McGibbon$^{1}$, Brian Henn$^{1}$, Oliver Watt-Meyer$^{1}$ \thanks{\texttt{oliverwm@allenai.org}}\\
  $^{1}$Allen Institute for Artificial Intelligence (Ai2), Seattle, USA\\
  $^{2}$Department of Electrical and Computer Engineering, University of Washington, Seattle, USA
}

\begin{document}

\maketitle

\begin{abstract}

In the recent AI Model Intercomparison Project (AIMIP) benchmark, the ACE2.1-ERA5 emulator was shown to underestimate the true near-surface warming trend both during the training period and the out-of-sample extrapolation period. In this work, we present M-ACE2S (Masked-ACE2S), a modified version of the stochastic ACE2S model, which uses variable masking during training and reduces the reliance of M-ACE2S on any single input channel. We show that M-ACE2S improves long-term near-surface air temperature prediction. We also present a novel data normalization strategy called ``global mean removal'', which, combined with masking, further improves the near-surface temperature forecasts. We show that M-ACE2S outperforms the ACE2S baseline on in-sample data for ERA5 while improving training stability and generalizability of near-surface temperature to unseen forcing when paired with global mean removal.

\end{abstract}

\section{Introduction}
\label{sec:introduction}

The recent AI Model Intercomparison Project (AIMIP) \cite{henn2026aimip} systematically evaluates AI-based climate emulators under a common experiment protocol. Models that participated in AIMIP were restricted from using CO$_2$ as an input and each model was evaluated on its ability to reproduce the near-surface warming trend solely from the prescribed sea ice and sea surface temperature. AIMIP follows the Atmospheric Model Intercomparison Project (AMIP) design \cite{gates1999amip}, in which sea surface temperature (SST) and sea ice are prescribed from observations. In this configuration the near-surface air over the ocean is tied to the prescribed SST by turbulent surface fluxes, and ocean warming is communicated to land through warmer and moister advected air, increased downward longwave radiation, and the lapse-rate response of the free troposphere \cite{compo2009oceanic,dong2009understanding}. Consistent with this, the near-surface temperature of a physics-based climate model warms when its prescribed ocean warms even with CO$_2$ held at pre-industrial levels, as in the amip-piForcing experiments that reproduce the historical global-mean warming \cite{gregory2016variation,andrews2018accounting}. A key finding is that ACE2.1-ERA5 \cite{henn2026aimip}, a modified version of the deterministic ACE2 \cite{watt2025ace2}, does not properly learn the historical warming trend of ERA5's near-surface air temperature from the prescribed SST. The previous ACE2-ERA5 model, which received CO$_2$ as an input, was able to reproduce this trend \cite{henn2026aimip,watt2025ace2}. NeuralGCM \cite{kochkov2024neural}, ArchesWeatherGen \cite{couairon2024archesweather}, and DLESyM \cite{cresswell2025deep} capture most of it under the CO$_2$-excluded AIMIP protocol \cite{henn2026aimip}.

ACE2 models that did receive CO$_2$ appear to rely on its monotonic rise more than on sea surface temperature. As shown in \cite{watt2025ace2} and \cite{clark2026disentangling}, when ACE2-SHiELD was given fixed 1979 CO$_2$ concentration the model fails to warm at the expected rate. ACE2-SOM likewise unphysically predicts that shortwave and surface upward longwave fluxes change when CO$_2$ varies while the atmospheric and surface states are held fixed; the authors attribute this to a lack of exposure to independent combinations of CO$_2$ and atmospheric states during training \cite{clark2025ace2}. Because CO$_2$ is a monotonically increasing field (at least for its annual mean), an emulator trained on the historical record can use it as a clock rather than as a forcing \cite{henn2026aimip}. Sea surface temperature carries a weaker version of the same risk \cite{henn2026aimip,brenowitz2025climate}, since its trend is small relative to its seasonal and interannual variability. Withholding CO$_2$ under AIMIP therefore removes the CO$_2$ shortcut and tests whether an emulator can reproduce the trend from the prescribed sea surface temperature and sea ice alone, the only inputs that carry a trend.

This paper addresses the difficulty ACE2.1-ERA5 has in learning the warming trend without CO$_2$, through a data augmentation approach of variable masking during training, which requires the model to predict each variable from the remaining inputs rather than from any single channel. Here, we adopt the AIMIP protocol as our evaluation framework and apply a masked-training approach to the stochastic version of ACE2, called ACE2S \cite{perkins2025hiro,clark2026disentangling}. We focus on the ACE family of models because it is an actively developed and maintained base for climate applications: ACE2 reproduces historical variability and trends on ERA5 \cite{watt2025ace2}, and its variants emulate slab-ocean climate sensitivity to CO$_2$ \cite{clark2025ace2}, fully coupled ocean--atmosphere climates \cite{wu2026stochastic}, and km-scale precipitation through downscaling \cite{perkins2025hiro}, so improvements to ACE2S carry over to these uses. We call this model M-ACE2S. We show that our masking approach stabilizes model training, improves in-sample forecast skill, and, when paired with global mean removal, improves generalization to out-of-sample climate data. 

\section{Methods}
\label{sec:methods}

\subsection{Dataset}

Many ML climate models are trained on the ERA5 reanalysis dataset \cite{Hersbach2020} because it is the best representation of historical atmospheric conditions that is publicly available today. Following the AIMIP protocol \cite{henn2026aimip}, we train on ERA5 from 1979 through 2008, validate on 2009 through 2014, and test on 2015 onward. The AIMIP protocol excludes CO$_2$ and all other anthropogenic radiative forcers as model inputs because they have been shown to be potential clock-like variables in ACE that leads to overfitting. Sea surface temperature, sea ice, and the seasonal cycle of insolation are therefore the only historically varying forcings. We use a 1\textdegree\ Gaussian grid ($180 \times 360$) for the training data. We coarsen the vertical coordinate to 8 layers while conserving moisture and energy. Preprocessing follows ACE2 \cite{watt2025ace2} and ACE2S \cite{perkins2025hiro}; one ERA5-specific deviation is noted in \autoref{appendix:data_preprocessing}. The 2009--2014 validation period is used for the rollout-based checkpoint selection of \autoref{appendix:checkpoint_selection}; here, the single-step validation loss that we report alongside the training loss is computed on the years 1994 and 2014 (instead of 2009--2014), of which only 2014 lies outside the training window.

\subsection{Training}
\label{subsec:training}

\paragraph{M-ACE2S} 
M-ACE2S is a masked-training version of ACE2S. ACE2S was first presented in HiRO-ACE \cite{perkins2025hiro} and is the stochastic version of ACE2 \cite{watt2025ace2}. ACE2S uses the SFNO architecture \cite{bonev2023spherical} and is a nonlinear function $[P_{t+1}, D_{t+1}] = f(P_t, F_t)$ that maps prognostic variables $P_t$ and forcing variables $F_t$ to the next-step prognostic variables $P_{t+1}$ and the diagnostic variables $D_{t+1}$. We use uniformly spaced time steps of 6 hours for the training data ($\Delta t = 6\,\mathrm{hr}$). Network inputs and outputs use the same two-stage ``full-field'' / ``residual'' normalization as ACE \cite{watt2023ace} and ACE2 \cite{watt2025ace2}, with the residual rescaling $a_{\mathrm{res}}$ applied to prognostic variables in the loss only. Following ACE2S \cite{perkins2025hiro} and ACE2 \cite{watt2025ace2}, a physical corrector that enforces non-negativity for variables where negative numbers do not make physical sense (such as precipitation and moisture), as well as conservation of global dry air mass and column moisture at each 6-hour step, is applied inside the optimized model but before taking the loss. Like ACE2S \cite{perkins2025hiro}, the 1\textdegree\ models are trained in two stages: the first is a 150-epoch single-step objective and the second is a 20-epoch multi-step objective (see \autoref{appendix:multistep_finetuning}). Details about the loss function are available in \autoref{appendix:loss_function}. Hyperparameters are described in \autoref{appendix:hyperparameters}. The model has about $8.0 \times 10^8$ trainable parameters.

\paragraph{Global Mean Removal}
We found that ACE2S has difficulty representing the global mean warming trend through the training period, despite the availability of sea surface temperature forcing, which is highly correlated with this trend. This is made more immediately apparent if near-surface air temperature is represented as its departure from the global mean surface temperature, instead of in absolute terms. In such a basis the model's predictions are clearly out of distribution when temperature trends do not follow the surface temperature trend.

We take advantage of this by applying global mean removal (GMR) before normalization: the per-sample global spatial mean of the surface temperature is subtracted from all eleven temperature fields (surface, 2\,m, 850\,hPa, and the eight air-temperature layers), so the network sees and directly predicts the departure of each field from that mean rather than its absolute magnitude at each level. The removed mean is appended as an extra normalized input channel and restored on the output fields after denormalization, making GMR an invertible reparameterization.

\paragraph{Masked Training}
Our masking approach is inspired by the Moirai 2.0 \cite{liu2025moirai} and PatchTST \cite{nie2022time} foundation models. During training, we draw a count $k$ uniformly from $\{0, 1, \ldots, 20\}$ and then mask $k$ input channels drawn without replacement from the 43 input variables, plus the appended global-mean channel when global mean removal is enabled (44 maskable channels in that case). Each GPU draws one mask per step, shared by all samples in its local batch. We mask the input variables $P_t$ and $F_t$ with a multiplicative mask $M_t$ that is 0 for the masked variables and 1 for the unmasked variables, so a variable $v$ obtains value $0$ when masked and $v$ when unmasked. Since inputs are normalized by per-variable global statistics, a masked variable appears as its training-set global mean. We further append a dense binary masking channel for each input channel to help the model distinguish between masked and unmasked variables, doubling the width of the network input to 88 channels (86 without global mean removal). We then train the model to predict all of $P_{t+1}$ and $D_{t+1}$, including the channels that were masked out of the input, given the masked input variables $P_t$, $F_t$, and the masking channels $M_t$. At inference no channel is masked, so $M_t = \mathbf{1}$ and every mask indicator channel is one. We provide ablations of the masking approach in \autoref{appendix:masking_ablations}.

\subsection{Evaluation}
\label{subsec:evaluation}
We select the reported checkpoint by the channel-mean global root-mean-square (RMS) time-mean bias (normalized units) over eight 5-year validation rollouts started across 2009 (AIMIP rollouts). The 46-year climate rollouts are generated from the best AIMIP-rollout checkpoint of the 20 fine-tuning epochs, which was itself initialized from the lowest-validation-loss checkpoint of the 150 pre-training epochs. More details for checkpoint selection can be found in \autoref{appendix:checkpoint_selection}.

We evaluate M-ACE2S and the unmasked ACE2S baseline \cite{perkins2025hiro} with autoregressive inference runs, generated from a 1\textdegree, 8-layer, 6-hour resolution split of the ERA5 dataset spanning 1979 through 2024. Inferences use eight initial conditions, spaced roughly six weeks apart for the AIMIP rollouts and one day apart for the 46-year climate rollouts. Each run free-runs the model forward from its start date without any nudging toward observations. Our results use the mean over the resulting eight-member initial-condition ensemble for each model. All results in the main text are for models trained with global mean removal; the corresponding models trained without it are reported in \autoref{appendix:additional_1deg_results}. The full configuration file of every trained model, including the ablations of \autoref{appendix:masking_ablations}, is released with the code (see \hyperref[fn:code_availability]{code availability}); \autoref{appendix:configurations} summarizes them.

\section{Results}
\label{sec:results}
\subsection{46-year Climate Rollouts}

We evaluate the models with 46-year climate rollouts on the ERA5 dataset. Of interest is the long-term trend of the near-surface air temperature $T_7$, which was a weakness of ACE2.1-ERA5 in the first part of the AIMIP benchmark \cite{henn2026aimip}.

\begin{figure}[ht]
  \centering
\includegraphics[width=0.65\textwidth]{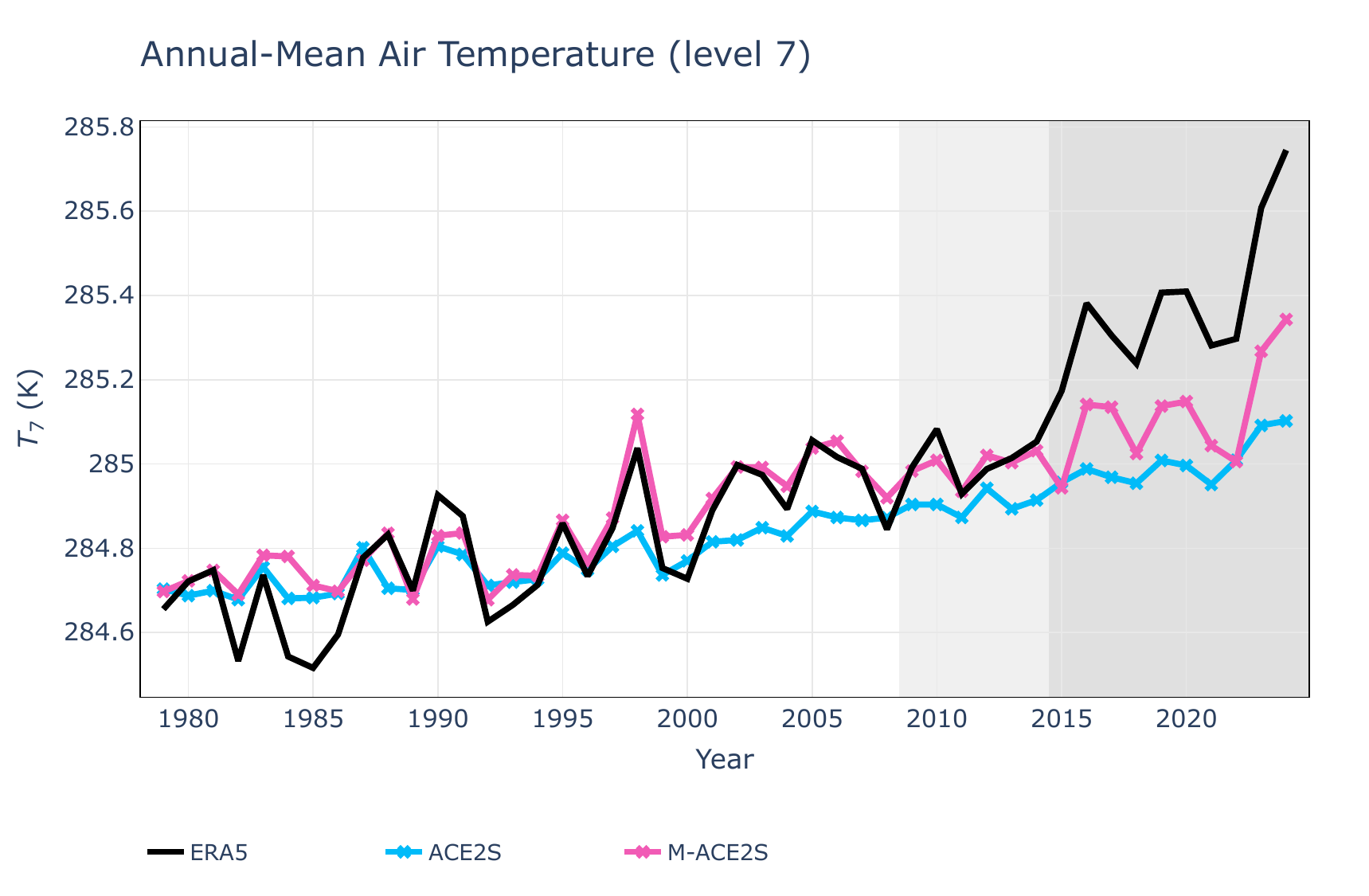}
  \caption{Global mean timeseries of near-surface air temperature $T_7$ for ERA5 (black), M-ACE2S (pink), and ACE2S (blue). The plotted line is the mean of the eight-member initial-condition ensemble for each model. Light and dark shading mark the AIMIP validation (2009--2014) and test (2015--2024) periods.}
  \label{fig:46yr_era5_T7_1deg}
\end{figure}

\autoref{fig:46yr_era5_T7_1deg} shows long-time forecasts of $T_7$ on ERA5 for M-ACE2S and ACE2S. We observe that M-ACE2S improves on in-sample (1979--2008) $T_7$ forecasts. Both models underestimate the out-of-sample (2015--2024) warming, but M-ACE2S underestimates it less. 

\autoref{table:t7_trend_1deg} shows the RMSE, bias, and linear-trend error of $T_7$ on ERA5 for both models. Each underestimates the test-period warming trend. Masking lowers the RMSE and trend error in both periods, reducing the test-period trend error from $-0.28$ to $-0.18$\,K\,decade$^{-1}$ and the test-period RMSE from 0.40 to 0.27\,K. Corresponding figures and tables for total water path $TWP$ are in \autoref{appendix:additional_1deg_results}.

\begin{table}[ht]
  \caption{Errors of the annual global-mean $T_7$ for ERA5 over the AIMIP training (1979--2008) and test (2015--2024) periods. RMSE: root-mean-square of model minus ERA5 over the years of the period. Bias: mean of model minus ERA5. Trend error: model minus ERA5 least-squares linear trend. Bold marks the smaller magnitude of each pair.}
  \label{table:t7_trend_1deg}
  \centering
  \small
  \begin{tabular}{lcccccc}
    \toprule
     & \multicolumn{2}{c}{RMSE (K)} & \multicolumn{2}{c}{Bias (K)} & \multicolumn{2}{c}{Trend error (K\,decade$^{-1}$)} \\
    \cmidrule(lr){2-3} \cmidrule(lr){4-5} \cmidrule(lr){6-7}
     & Train & Test & Train & Test & Train & Test \\
    \midrule
    ACE2S (GMR on) & 0.10 & 0.40 & \textbf{$-$0.02} & $-$0.38 & $-$0.061 & $-$0.279 \\
    M-ACE2S (GMR on) & \textbf{0.08} & \textbf{0.27} & 0.04 & \textbf{$-$0.27} & \textbf{$-$0.015} & \textbf{$-$0.177} \\
    \bottomrule
  \end{tabular}
\end{table}

\begin{figure}[ht]
  \centering
  \includegraphics[width=0.95\textwidth]{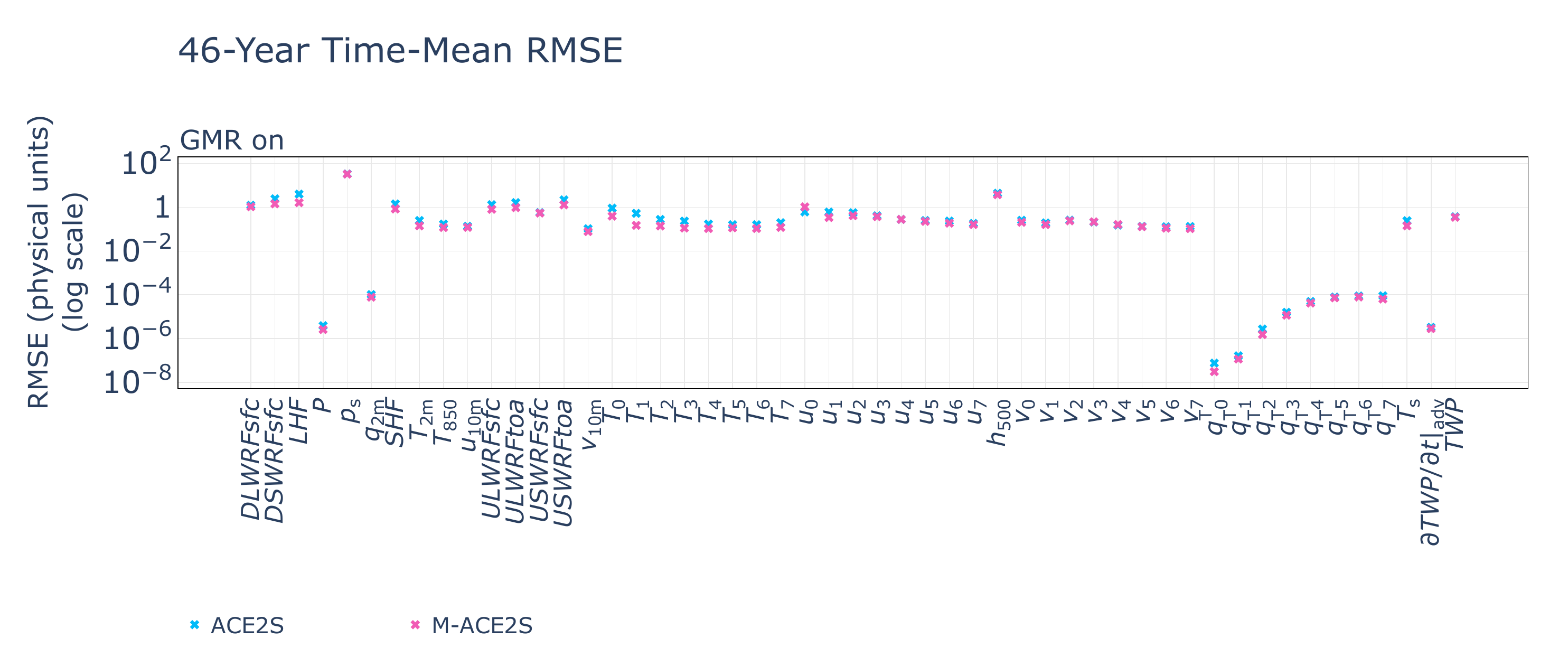}
  \caption{The time-mean RMSE of the model's 46 year rollout on ERA5 for all output variables.}
  \label{fig:46yr_era5_time_mean_rmse_gmron_1deg} 
\end{figure}

We also observe the effect of masking on the time-mean RMSE on ERA5 for all output variables in \autoref{fig:46yr_era5_time_mean_rmse_gmron_1deg}. We see that for 47 of the 51 fields, the masked model has lower time-mean RMSE than the unmasked model.

\begin{figure}[ht]
  \centering
  \includegraphics[width=0.45\textwidth]{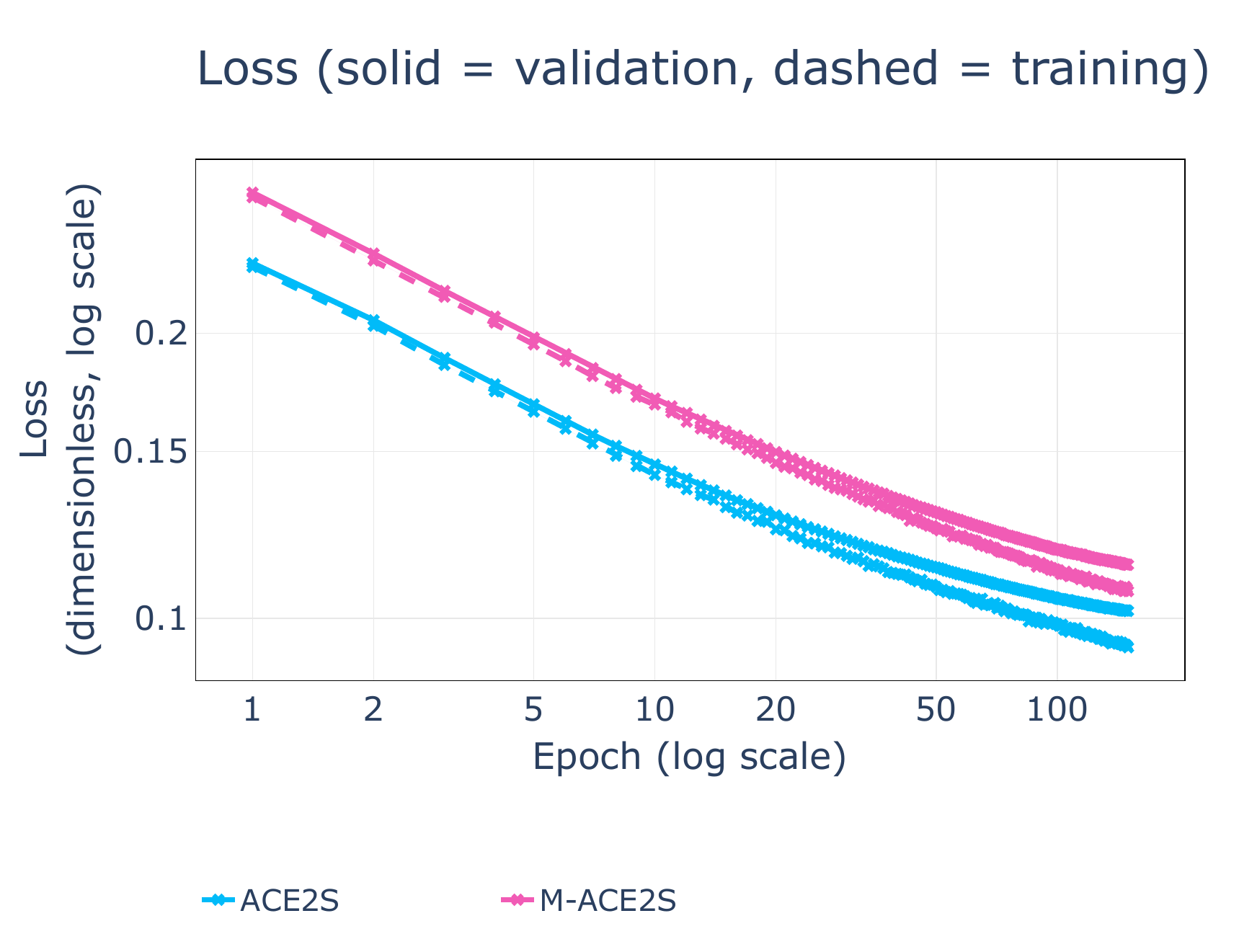}
  \includegraphics[width=0.45\textwidth]{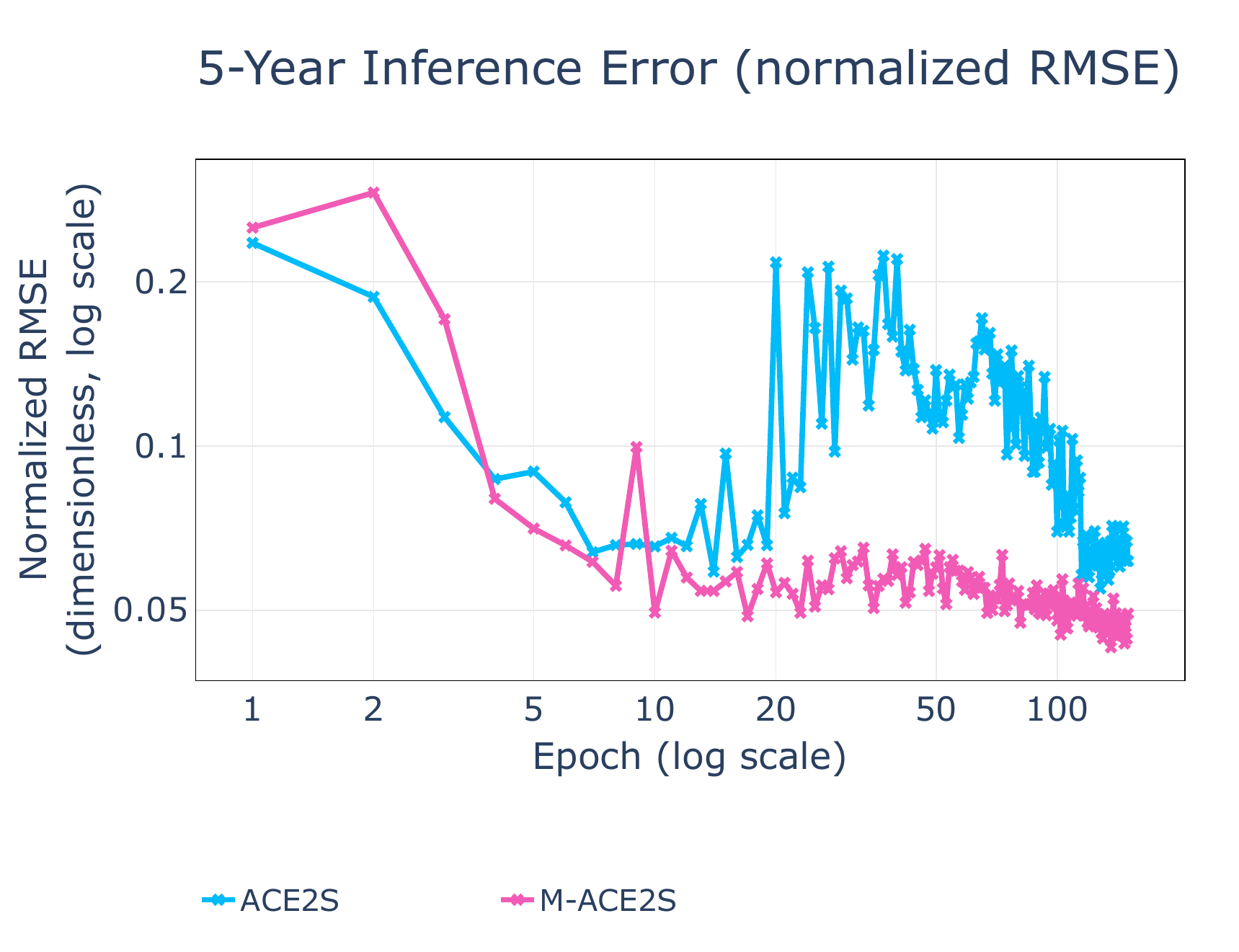}
  \caption{Training (dashed) and validation (solid) loss (left) and AIMIP checkpoint time-mean normalized RMSE averaged over channels (right) during training for M-ACE2S (pink) and ACE2S (blue).}
  \label{fig:46yr_era5_training_metrics_1deg_gmron}
\end{figure}

\autoref{fig:46yr_era5_training_metrics_1deg_gmron} shows the training dynamics of the two models. We note that masking increases the training and validation loss throughout training, but reduces the difference between the two. These two observations have a common explanation. Since masking is a regularizer, it reduces overfitting of the model to the training data, closing the gap between the training and validation loss. However, since masking is a harder objective to train on by asking the model to predict a variable that was not passed in as an input, the two losses are overall higher. Furthermore, the masked models achieve overall lower 5-year inference errors (calculated as the time-mean RMSE over the AIMIP validation years of 2009--2014), which demonstrates that the model is obtaining greater forecasting capabilities due to masked training. Finally, masking appears to alleviate a long-standing issue observed in the training of ACE where long training does not always exhibit lower climate time-mean biases.

\section{Conclusions and Future Work}
\label{sec:conclusions}

In this paper, we explore the addition of data augmentation to the training pipeline of the stochastic version of the ACE model (M-ACE2S) using the AIMIP protocol with the goal of reducing overfitting and reliance on any single input channel. We use a simple channel masking approach in conjunction with global mean removal from temperature fields, which helps M-ACE2S generalize its near-surface temperature forecasts to out-of-sample SSTs, track the ground-truth ERA5 trend more closely within the training period of 1979--2008, and reduce the time-mean RMSE for most variables relative to the unmasked ACE2S baseline. The training dynamics demonstrate that masking acts as a regularizer by reducing the gap between validation and training loss, which lowers 5-year inference errors on the 2009--2014 validation period from roughly epoch 20 onward. Beyond AIMIP, masking and global mean removal move ACE2S toward trends learned from the prescribed boundary forcing rather than memorized from the training period. Such trends are a prerequisite for the uses that motivate fast emulators in the first place: large initial-condition ensembles that estimate internal variability and the likelihood of extreme events \cite{watt2025ace2}, coupled atmosphere-ocean simulations in which ocean heat uptake must be simulated rather than prescribed \cite{wu2026stochastic}, and km-scale downscaling of precipitation for extreme-event risk assessment \cite{perkins2025hiro}. Masking does not improve every field: total water path drifts further from ERA5 over the test period (\autoref{appendix:additional_1deg_results}), so a masking strategy that targets specific variables remains open.
\begingroup
\renewcommand{\thefootnote}{}
\footnote{\label{fn:code_availability}Code availability: training code is available at \url{https://github.com/ai2cm/ace}. Code to reproduce the plots and tables, along with the training and inference configuration file of every model in this paper, is available at \url{https://github.com/ai2cm/m-ace2s-paper}.}
\addtocounter{footnote}{-1}
\endgroup

Opportunities remain to improve the masking strategy. Our initial testing used a Bernoulli masking strategy, where each variable was assigned a fixed probability of being masked during training (ablation shown in \autoref{appendix:masking_ablations}). Switching to the uniform masking strategy we presented here, we found better results on long-time forecasts. We leave additional masking strategies for future work but outline some potential avenues of research here. First, a climate-informed masking strategy could be developed to mask out variables that the model is over-weighting when generating climate forecasts. Instead of randomly selecting which variables to mask, during training the model could select variables that would benefit from masking dynamically. Another direction is to use spatial masking, where only portions of a variable's spatial field are masked during training; this would, however, require architectures that do not have an SFNO backbone, in order to avoid spectral ringing. Finally, a thorough investigation of post-mask training without masking could be conducted. Currently, masking occurs only during training, but we evaluate the model's forecasts without a mask; these are disjoint regimes.

\vspace{-1mm}
\begin{ack}
\vspace{-2mm}
Ai2 acknowledges support from the Fund for Science and Technology. We also acknowledge ECMWF for generating and making available the ERA5 dataset.
\end{ack}

\bibliographystyle{unsrtnat}

\newpage
\appendix
\raggedbottom

\section{Data Preprocessing}
\label{appendix:data_preprocessing}
We follow the same pre-processing pipeline used for ACE \cite{watt2023ace} and ACE2 \cite{watt2025ace2}, which is also the pipeline used to build the ERA5 pretraining dataset for ACE2S \cite{perkins2025hiro}; see those works for details.

ERA5 is assembled from several production streams, and the fields carry a small discontinuity at each stream-to-stream boundary. Four such boundaries fall inside our training window (between 1986-03-31 and 1986-04-01, between 1993-07-31 and 1993-08-01, between 1993-12-31 and 1994-01-01, and between 1999-12-31 and 2000-01-01). We split the training dataset at each of them, so no training sample pairs a time step on one side of a stitch with a time step on the other and the model is never asked to predict across the jump, which would otherwise present a coherent bias step as a one-step target.

\section{Loss Function}
\label{appendix:loss_function}
M-ACE2S is trained with the same stochastic ensemble loss used for the atmospheric component of SamudrACE \cite{wu2026stochastic}. M-ACE2S differs in that it is atmosphere-only, so the ocean term, the land masking of ocean fields, and the relative weighting between components are absent, and the loss reduces to the single-component form. The loss is only computed during the last forward step for stage 1 and stage 2, as outlined in \autoref{appendix:multistep_finetuning}. Masking is applied to the network inputs only (\autoref{appendix:masking_ablations}); every output channel is scored at every training step regardless of which inputs were masked, so the masked and unmasked models optimize exactly the same objective, differing only in the model inputs.

\section{Data Variables}
\label{appendix:data_variables}

The training datasets use the same variables as ACE2.1-ERA5, which are listed in \cite{henn2026aimip}.

\section{Hyperparameters}
\label{appendix:hyperparameters}

We use the training hyperparameters listed in \autoref{table:optimizationparams} and the model hyperparameters listed in \autoref{table:sfnoparams} for the training of M-ACE2S. Both tables describe the 1\textdegree\ 6-hourly models of the main text; \autoref{appendix:additional_4deg_results} lists the few settings that differ for the 4\textdegree\ models. Values not specific to masked training follow ACE2S \cite{perkins2025hiro}, which in turn inherits its architecture hyperparameters (embedding size, per-variable weighting, number of SFNO blocks) from ACE2 \cite{watt2025ace2}.

\begin{table}[H]
  \caption{NC-SFNO hyperparameters. Names correspond to the \texttt{NoiseConditionedSFNO} builder of the ACE codebase, whose call signature follows the \texttt{SphericalFourierNeuralOperatorNet} of \url{https://github.com/ai2cm/ace}. All configuration options not listed here are left at their defaults; in particular the spectral filter is neither grouped nor band-limited (\texttt{filter\char`_num\char`_groups} $=1$, \texttt{spectral\char`_ratio} $=1$), so each of the 8 blocks carries a full $180 \times 512 \times 512$ complex per-mode weight, which together with the $512 \times 180 \times 360$ positional embedding accounts for most of the $8.0 \times 10^8$ parameters.}
  \label{table:sfnoparams}
  \centering
  \begin{tabular}{ll}
    \toprule
    Name     & Value \\
    \midrule
    \texttt{embed\char`_dim}                  & 512       \\
    \texttt{noise\char`_embed\char`_dim}      & 32        \\
    \texttt{noise\char`_type}                 & isotropic \\
    \texttt{filter\char`_type}                & linear    \\
    \texttt{affine\char`_norms}               & true      \\
    \texttt{normalize\char`_big\char`_skip}   & true      \\
    \texttt{use\char`_mlp}                    & true      \\
    \texttt{mlp\char`_ratio}                  & 2.0       \\
    \texttt{num\char`_layers}                 & 8         \\
    \texttt{operator\char`_type}              & dhconv    \\
    \texttt{clip\char`_latent\char`_global\char`_means} & false \\
    \bottomrule
  \end{tabular}
\end{table}

\begin{table}[H]
  \caption{Optimization hyperparameters for the 1\textdegree\ 6-hourly models. The batch size is global: it is divided across the 8 GPUs of a training job, so each GPU sees one sample per step.}
  \label{table:optimizationparams}
  \centering
  \begin{tabular}{ll}
    \toprule
    Name     & Value \\
    \midrule
    Optimizer                              & FusedAdam  \\
    Initial learning rate                  & $1 \times 10^{-4}$    \\
    Learning rate schedule                 & Constant   \\
    Weight decay                           & 0.01       \\
    Number of epochs                       & 150        \\
    Global batch size                      & 8          \\
    GPUs per training job                  & 8          \\
    Gradient accumulation                  & Enabled    \\
    Automatic mixed precision              & Disabled   \\
    Forward steps per training step        & 1          \\
    Ensemble members per training step     & 2          \\
    Exponential moving average decay rate  & 0.999      \\
    Validation / inference weights         & EMA        \\
    \bottomrule
  \end{tabular}
\end{table}

\section{Trained Configurations}
\label{appendix:configurations}

\autoref{table:configurations} lists every trained configuration in this paper, grouped by experiment, with the data, normalization, masking, seeds, and training schedule of each, and the section in which its results appear.

\begin{table}[H]
  \caption{Trained configurations. See \hyperref[fn:code_availability]{code availability} for links to source code and raw training configurations. Data: horizontal resolution and time step of the ERA5 training data. Masking: uniform 0--$k$ masks a number of input channels drawn uniformly from $\{0, \ldots, k\}$, Bernoulli masks each channel with probability $p$; unless stated, one mask is shared across each GPU's batch and the model is told which channels are masked by appended channels. Seeds: number of trained seeds. Epochs: single-step (SS) and multi-step (MS) training (\autoref{appendix:multistep_finetuning}). Shown in: section or appendix with the results.}
  \label{table:configurations}
  \centering
  \small
  \setlength{\tabcolsep}{4pt}
  \begin{tabular}{lllp{3.3cm}lll}
    \toprule
    Experiment & Data & GMR & Masking & Seeds & Epochs & Shown in \\
    \midrule
    Main & 1\textdegree{}, 6 h & off, on & none, uniform 0--20 & 1 & 150 SS + 20 MS & \ref{sec:results}, \ref{appendix:additional_1deg_results} \\
    Main & 4\textdegree{}, 1 day & off, on & none, uniform 0--20 & 5 & 150 SS & \ref{appendix:additional_4deg_results} \\
    Mask distribution & 4\textdegree{}, 1 day & off, on & Bernoulli $p = 0.23$ & 5 & 150 SS & \ref{appendix:masking_ablations} \\
    Mask sharing & 4\textdegree{}, 1 day & on & uniform 0--10 / 0--20, per batch / sample & 1 & 150 SS & \ref{appendix:masking_ablations} \\
    Mask indicator & 4\textdegree{}, 1 day & on & uniform 0--20, channels / FiLM / both & 1 & 150 SS & \ref{appendix:masking_ablations} \\
    \bottomrule
  \end{tabular}
\end{table}

\section{Multi-Step Fine-Tuning}
\label{appendix:multistep_finetuning}

The 1\textdegree\ models are pre-trained on the single-step loss for the 150 epochs of \autoref{table:optimizationparams} and then fine-tuned for a further 20 epochs on the multi-step rollout scheme of ACE2S \cite{perkins2025hiro}. During multi-step fine-tuning, the model randomly selects from \{1, 2, 4, 12, 20\} forward steps with probabilities \{60\%, 20\%, 10\%, 5\%, 5\%\} respectively. Masking is applied only during the last forward step, which is also the only step being optimized.

\section{Checkpoint Selection}
\label{appendix:checkpoint_selection}

During training we periodically evaluate the model and report the checkpoint that best reproduces the long-term climatology, rather than the one with the lowest validation loss. ACE2S uses the same strategy on its X-SHiELD fine-tuning stage, selecting on the lowest channel-mean global RMS time-mean bias of multi-year rollouts \cite{perkins2025hiro}. Validation and AIMIP rollouts are performed with the exponential-moving-average (EMA) weights (decay $0.999$).

AIMIP rollouts are single-member, $\sim$5-year ($7300$ 6-hourly or $1825$ daily steps) free-runs initialized from eight start dates spaced roughly six weeks apart across 2009, inside the validation period. Checkpoints are scored by the channel-mean global RMS time-mean bias over these rollouts, with time means pooled over the eight runs and computed in full-field-normalized units.

Training and checkpoint selection for 1\textdegree\ models proceed in three stages:
\begin{enumerate}
  \item Single-step pre-training runs for $150$ epochs, and the checkpoint with the lowest validation loss is retained. Validation loss and the AIMIP rollouts are computed every epoch, so the candidate set is $\Theta_1 = \{\theta_e : e \in \{1, \ldots, 150\}\}$, where $\theta_e$ denotes the EMA weights at epoch $e$.
  \item Multi-step fine-tuning is initialized from that checkpoint and runs for a further $20$ epochs.
  \item After each of the $20$ fine-tuning epochs, $\Theta_2 = \{\theta_e : e \in \{1, \ldots, 20\}\}$, we compute this bias and report the checkpoint that minimizes it.
\end{enumerate}
The 4\textdegree\ models of \autoref{appendix:additional_4deg_results} have no second stage, so for them we report the checkpoint in $\Theta_1$ that minimizes this bias.

\section{Masked Training Ablations}
\label{appendix:masking_ablations}
We provide ablations of the masked training approach. The results here are models trained on 4\textdegree\ daily data, as opposed to 1\textdegree\ 6-hourly data presented in the main manuscript. In particular, we show that masking also helps on 4\textdegree\ data, motivate the method of providing the binary mask indicator to the model, and compare uniform masking to Bernoulli masking. These ablations were run at 4\textdegree\ only; the selected configuration was adopted unchanged at 1\textdegree, where it gives the gains of \autoref{sec:results}. The alternatives were not re-tested at 1\textdegree.

\begin{figure}[H]
  \centering
  \includegraphics[width=0.45\textwidth]{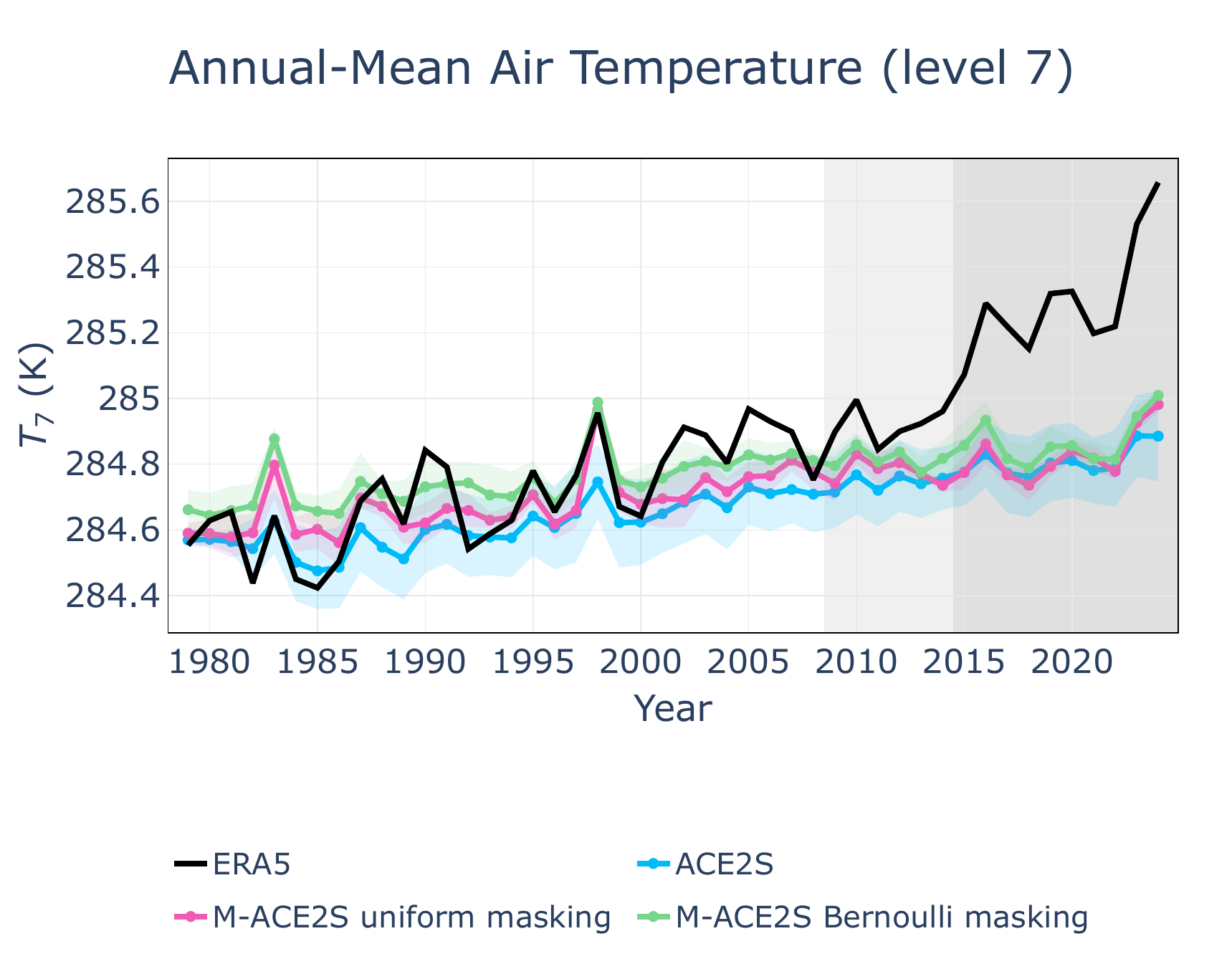}
  \includegraphics[width=0.45\textwidth]{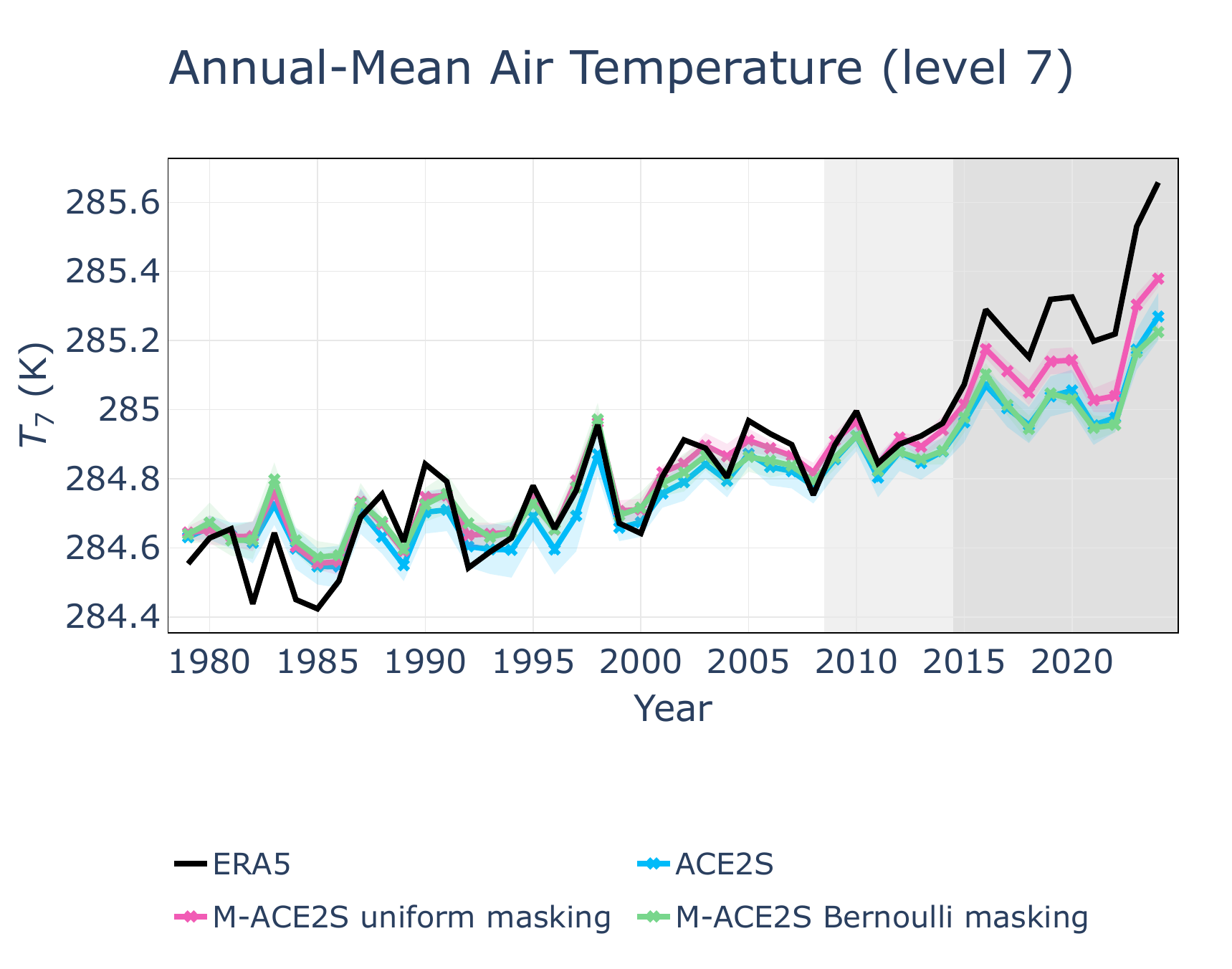}
  \caption{Global mean timeseries of near-surface air temperature $T_7$ for ERA5 (black), M-ACE2S uniform masking (pink), M-ACE2S Bernoulli masking (green), and ACE2S (blue). Shown are models trained without global mean removal (GMR) (left) and with global mean removal (GMR) (right). Mean and standard deviation are computed over the ensemble of 5 random seed runs. Results are for 4\textdegree\ daily models; this ablation was not run at 1\textdegree, so it has no 1\textdegree\ counterpart. Light and dark shading mark the AIMIP validation (2009--2014) and test (2015--2024) periods.}
  \label{fig:46yr_era5_T7_bern_4deg}
\end{figure}

\paragraph{Masking Distribution}
We show that masking with the Bernoulli distribution is less effective than uniform masking. In the main text, we sample masks by first drawing a count $k$ uniformly from $\{0, 1, \ldots, k_{\max}\}$ and then masking exactly $k$ variables, drawn without replacement, for each GPU's batch. In early experimentation, we used Bernoulli masking, where each variable was assigned some probability $p$ to be masked. Then, for each batch, each variable is masked with probability $p$ independent of other variables. The former we call uniform masking and the latter we call Bernoulli masking. \autoref{fig:46yr_era5_T7_bern_4deg} shows how uniform masking and Bernoulli masking compare to each other and to no masking on a 46-year forecast of $T_7$, our variable of interest. In particular, we select $k_{\max}=20$ and $p=0.2326$ so that in expectation, $10$ variables are masked regardless of masking method. Checkpoint selection is the same as in \autoref{appendix:checkpoint_selection}.

From the figure, we see that without global mean removal, Bernoulli and uniform masking work similarly, both reducing variance of the generated forecast across seeds. However, when global mean removal is being used, uniform masking outperforms Bernoulli masking. In fact, Bernoulli masking achieves similar performance to no masking.

\begin{figure}[H]
  \centering
  \includegraphics[width=0.45\textwidth]{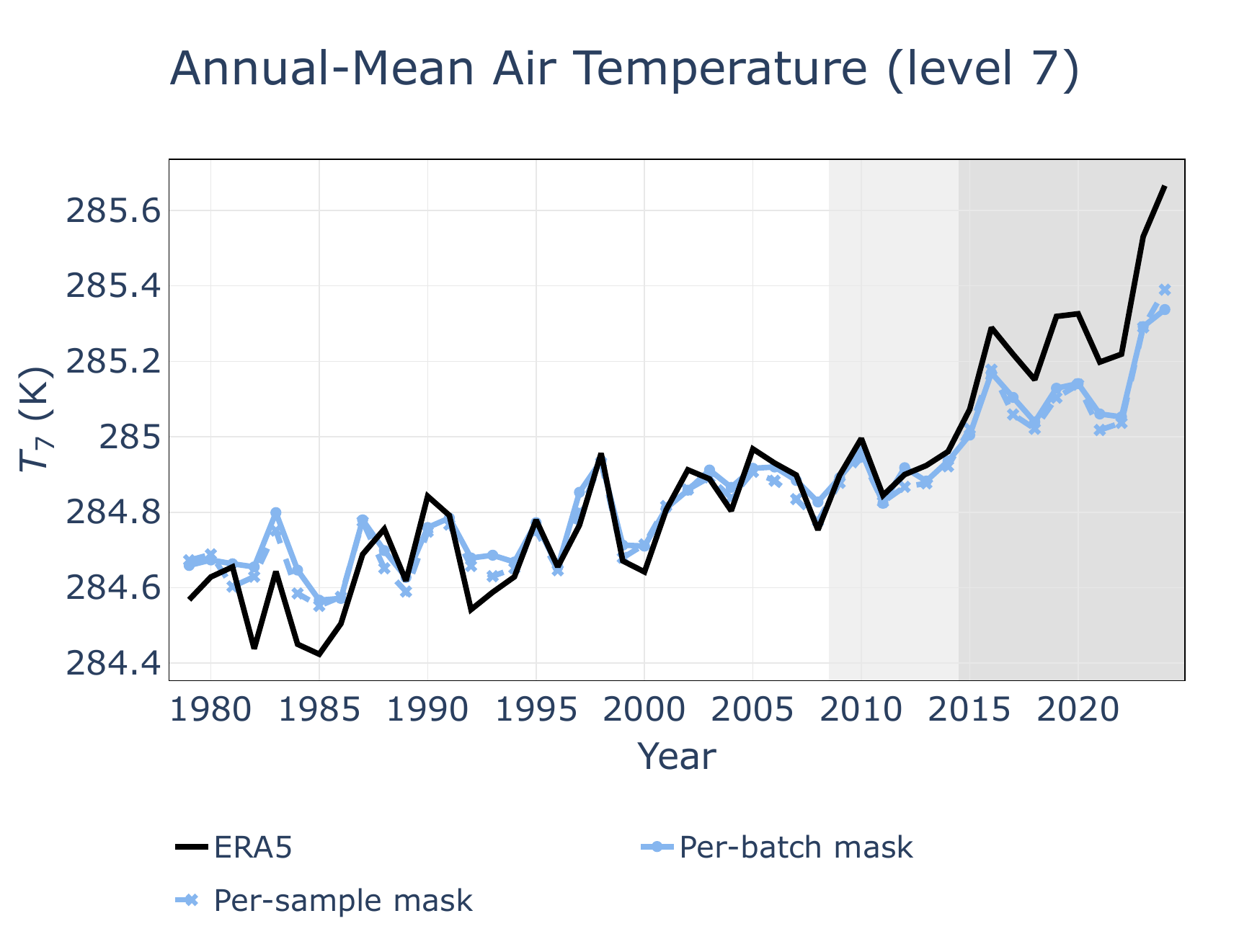}
  \includegraphics[width=0.45\textwidth]{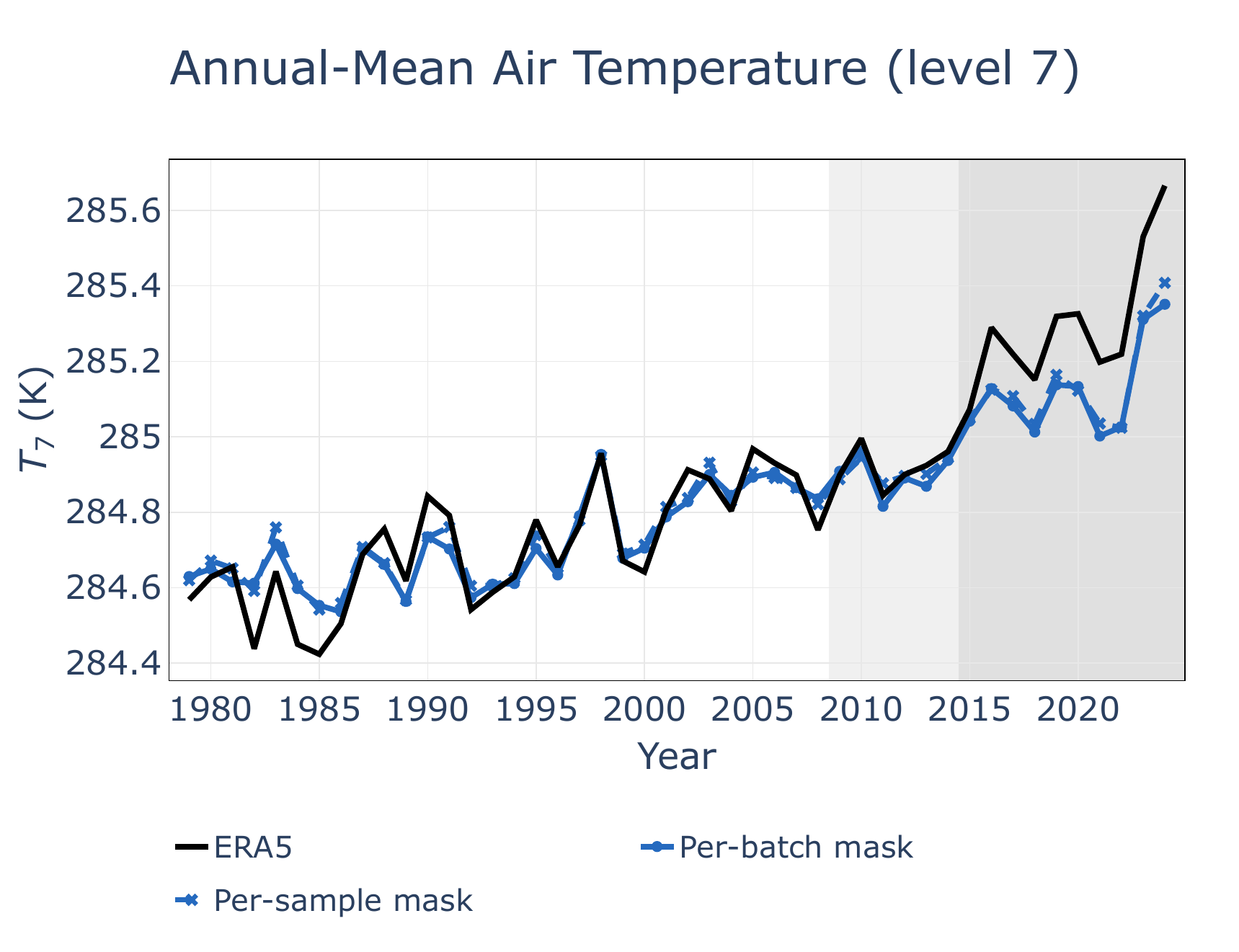}
  \caption{Global mean timeseries of near-surface air temperature $T_7$ for ERA5 (black) and for four M-ACE2S configurations, all trained with global mean removal (GMR): uniform 0--10 masking (left, light blue) and uniform 0--20 masking (right, dark blue), each with a single mask shared across each GPU's batch (``per-batch mask'', solid) or an independent mask drawn per sample (``per-sample mask'', dashed). One seed per config is shown. Results are for 4\textdegree\ daily models; this ablation was not run at 1\textdegree, so it has no 1\textdegree\ counterpart. Light and dark shading mark the AIMIP validation (2009--2014) and test (2015--2024) periods.}
  \label{fig:46yr_era5_T7_batch_comparison_4deg}
\end{figure}

\paragraph{Shared Mask Across Batch}
We show that sharing one mask across each GPU's batch is just as effective as drawing an independent mask for each sample. We show results in \autoref{fig:46yr_era5_T7_batch_comparison_4deg} comparing different methods of masking within a batch. We see that for two masking modes (uniform 0--10 and uniform 0--20), sharing one mask across the batch or masking each sample independently does not have an outsized effect on the 46-year rollout for $T_7$. All four configurations use the channel mask indicator and global mean removal of the main text, and differ only in the mask budget and in whether the mask is shared across the batch. We adopt the shared mask, drawn independently per GPU, because it is simpler and costs no accuracy.

\paragraph{Masking Indicator}

\begin{figure}[H]
  \centering
  \includegraphics[width=0.65\textwidth]{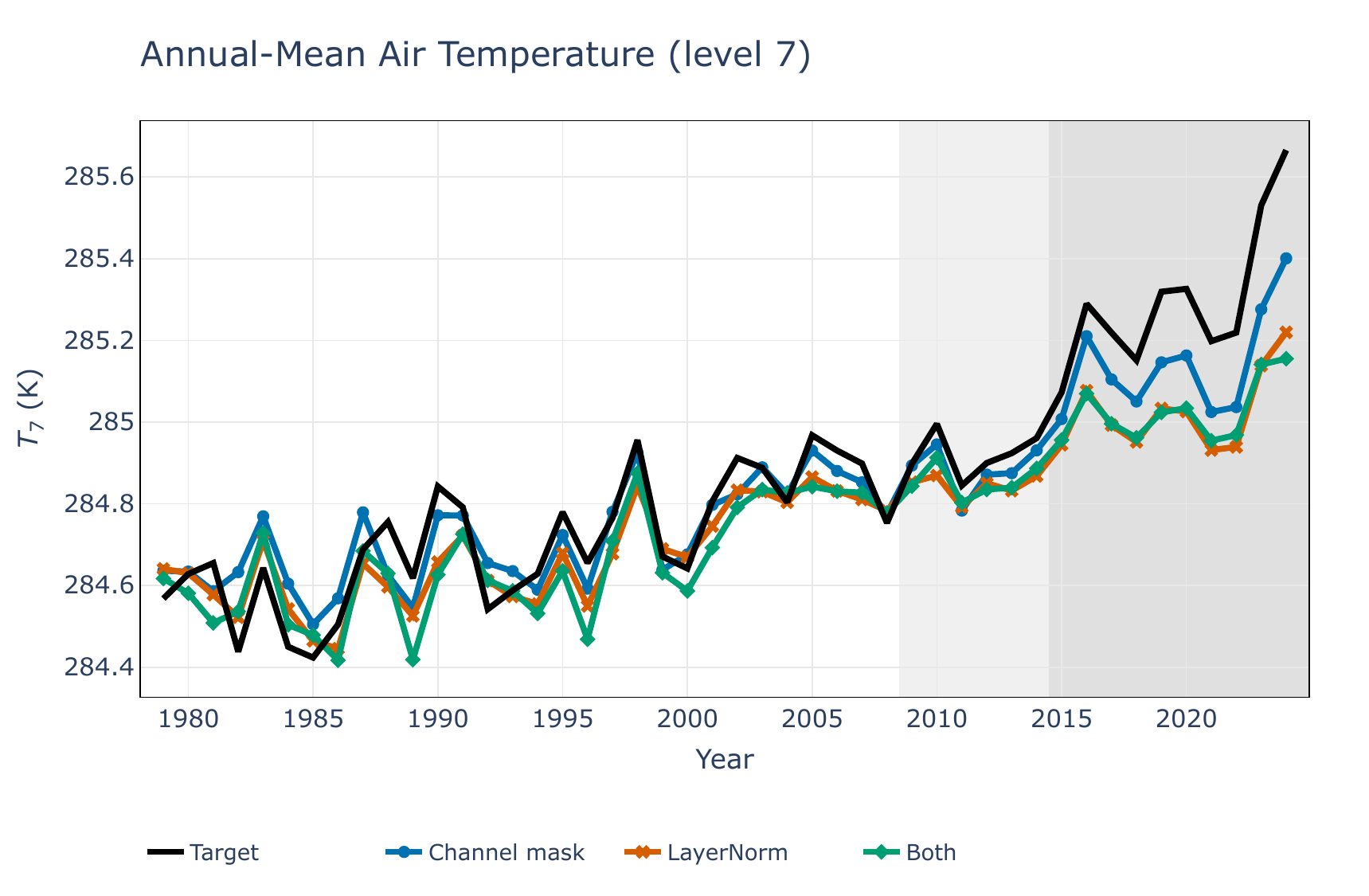}
  \caption{Global mean timeseries of near-surface air temperature $T_7$ for ERA5 (black, ``Target''), M-ACE2S with channel masking (blue, ``Channel mask''), M-ACE2S with FiLM masking (orange, ``LayerNorm''), and M-ACE2S with both FiLM and channel masking (green, ``Both''). Shown are models trained with global mean removal (GMR). One seed per config is shown. Results are for 4\textdegree\ daily models; this ablation was not run at 1\textdegree, so it has no 1\textdegree\ counterpart. Light and dark shading mark the AIMIP validation (2009--2014) and test (2015--2024) periods.}
  \label{fig:46yr_era5_T7_indicator_comparison_4deg}
\end{figure}

We show that passing the binary mask indicator as a separate channel to the model is more effective than passing it as a conditioning vector in the layer normalization. We show results in \autoref{fig:46yr_era5_T7_indicator_comparison_4deg} comparing different methods of providing variable masking information to the model. On a long rollout of $T_7$, channel masking (the method used in the paper) outperforms providing the mask to FiLM (conditional layernorm) and outperforms combining channel masking and FiLM. All three arms use uniform 0--20 masking with global mean removal, exactly as in the main text, and differ only in how the mask reaches the network: as appended binary input channels, as a FiLM conditioning vector, or as both.

\section{Additional 1\textdegree\ Results}
\label{appendix:additional_1deg_results}

\begin{figure}[ht]
  \centering
  \includegraphics[width=0.65\textwidth]{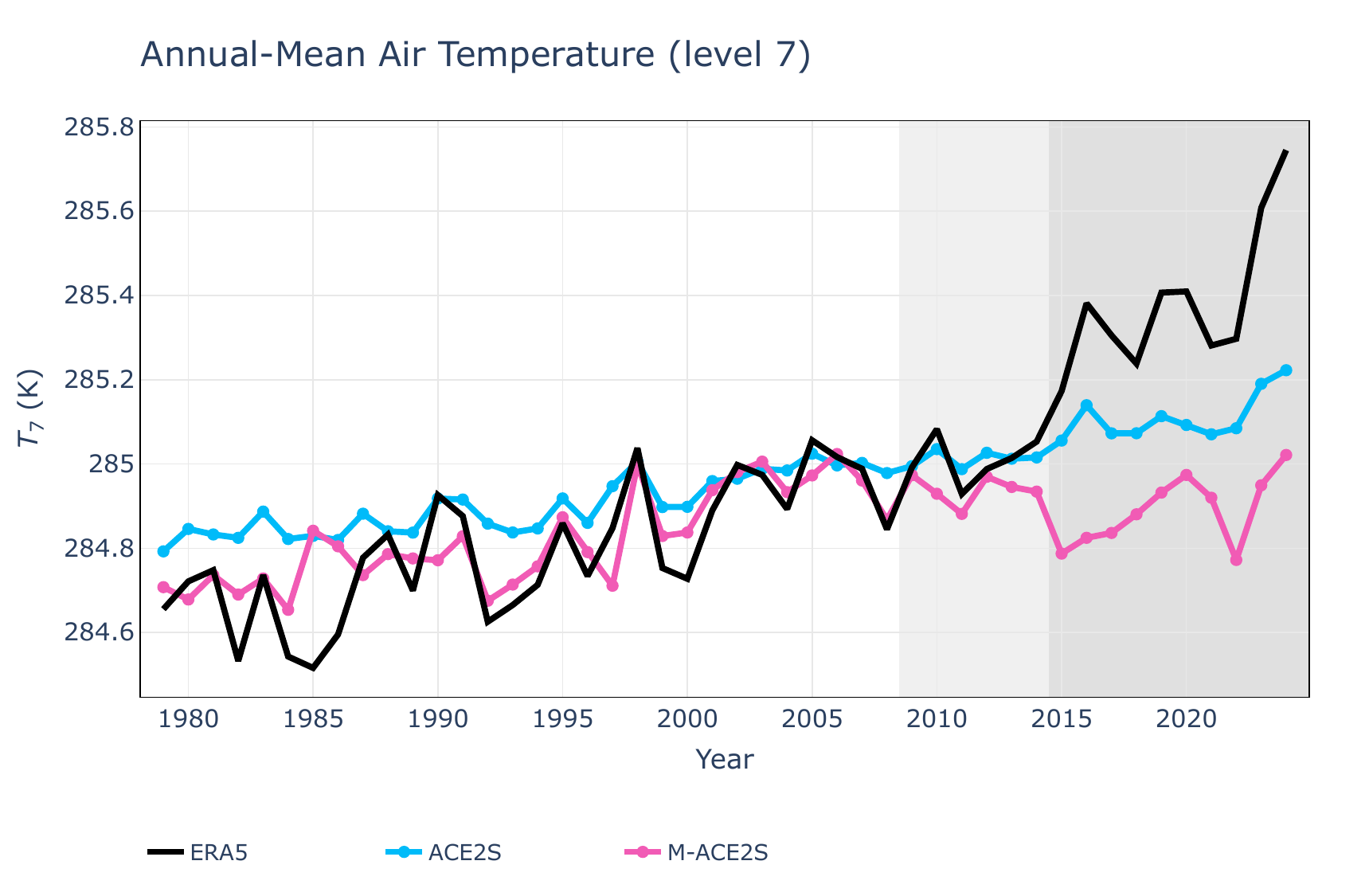}
  \caption{Global mean timeseries of near-surface air temperature $T_7$ for ERA5 (black), M-ACE2S (pink), and ACE2S (blue). Shown are models trained without global mean removal (GMR). The plotted line is the mean of the eight-member initial-condition ensemble for each model for the 46-year climate rollout. Results are for 1\textdegree\ 6-hourly models; see \autoref{fig:46yr_era5_T7_4deg} for the 4\textdegree\ counterpart. Light and dark shading mark the AIMIP validation (2009--2014) and test (2015--2024) periods.}
  \label{fig:46yr_era5_T7_gmroff_1deg}
\end{figure}

In \autoref{fig:46yr_era5_T7_gmroff_1deg} we show the 46-year $T_7$ forecast for the models trained without GMR.

\autoref{table:t7_trend_1deg_gmroff} scores these models as in \autoref{table:t7_trend_1deg}. Without GMR, masking lowers the trend error in both periods but raises the test-period RMSE from 0.30 to 0.51\,K, so masking alone does not improve the out-of-sample forecast; the two changes are complementary rather than interchangeable: masking regularizes, and GMR supplies the basis in which the regularized model extrapolates.

\begin{table}[H]
  \caption{Errors of the annual global-mean $T_7$ of the 46-year ERA5 rollouts of the 1\textdegree{} models trained \emph{without} global mean removal (GMR off), over the AIMIP training (1979--2008) and test (2015--2024) periods. Columns and conventions follow \autoref{table:t7_trend_1deg}.}
  \label{table:t7_trend_1deg_gmroff}
  \centering
  \small
  \begin{tabular}{lcccccc}
    \toprule
     & \multicolumn{2}{c}{RMSE (K)} & \multicolumn{2}{c}{Bias (K)} & \multicolumn{2}{c}{Trend error (K\,decade$^{-1}$)} \\
    \cmidrule(lr){2-3} \cmidrule(lr){4-5} \cmidrule(lr){6-7}
     & Train & Test & Train & Test & Train & Test \\
    \midrule
    ACE2S (GMR off) & 0.14 & \textbf{0.30} & 0.11 & \textbf{$-$0.27} & $-$0.058 & $-$0.299 \\
    M-ACE2S (GMR off) & \textbf{0.10} & 0.51 & \textbf{0.03} & $-$0.49 & \textbf{$-$0.025} & \textbf{$-$0.243} \\
    \bottomrule
  \end{tabular}
\end{table}

\begin{figure}[ht]
  \centering
  \includegraphics[width=0.45\textwidth]{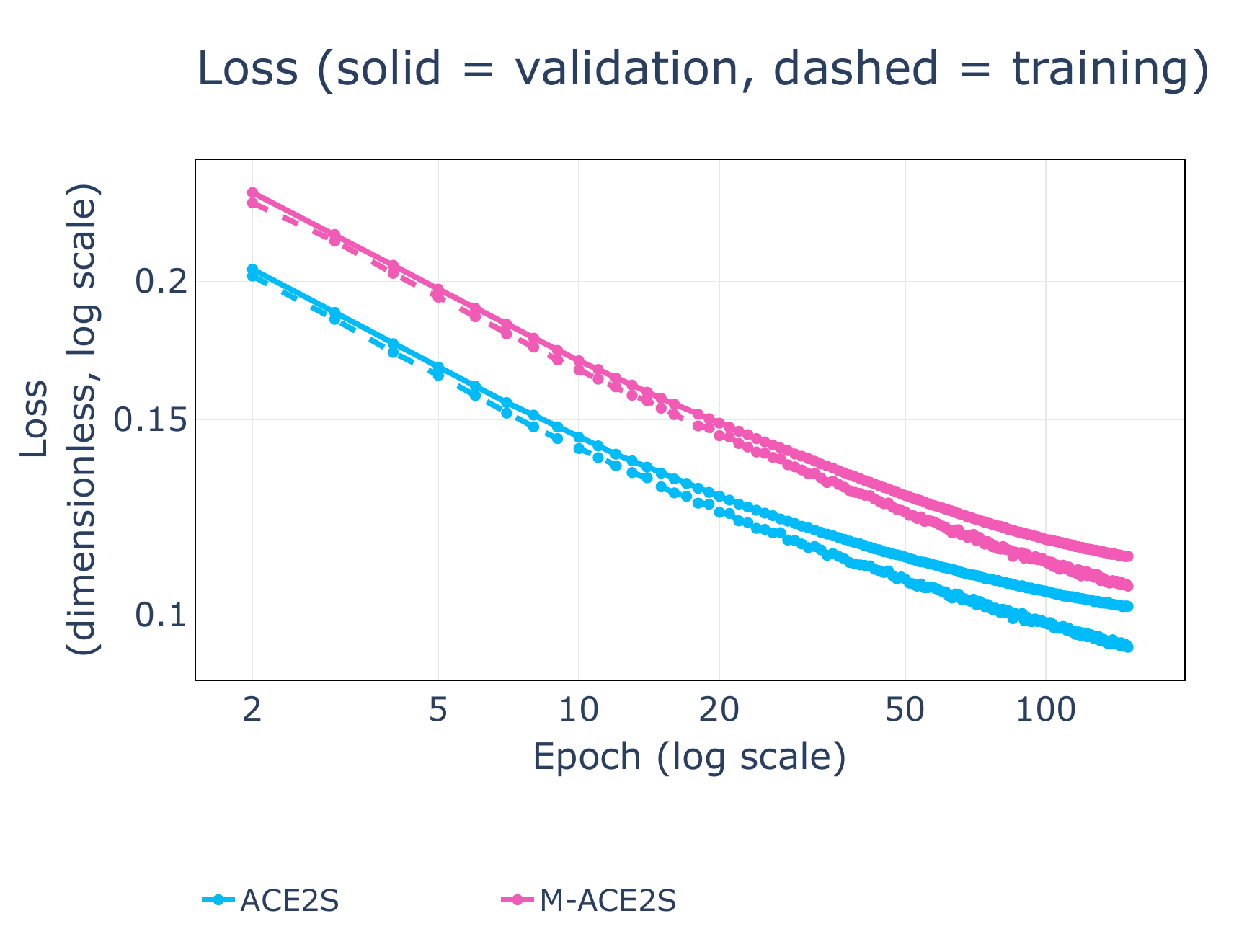}
  \includegraphics[width=0.45\textwidth]{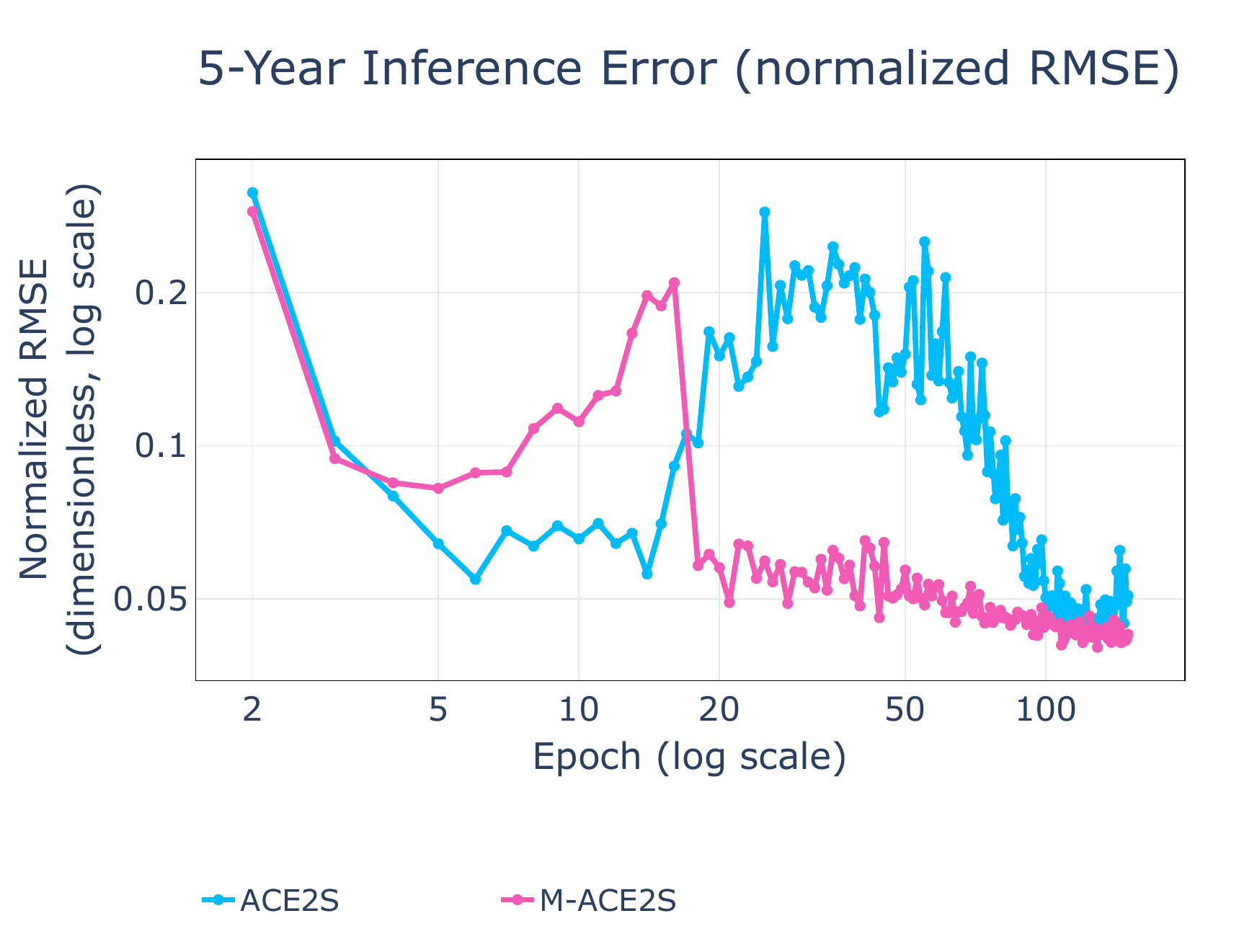}
  \caption{Training (dashed) and validation (solid) loss (left) and AIMIP checkpoint time-mean normalized RMSE averaged over channels (right) during training for M-ACE2S (pink) and ACE2S (blue) trained \emph{without} global mean removal (GMR off). Results are for the 1\textdegree\ 6-hourly models; see \autoref{fig:46yr_era5_training_metrics_4deg_gmroff} for the 4\textdegree\ counterpart.}
  \label{fig:46yr_era5_training_metrics_1deg_gmroff} 
\end{figure}

In \autoref{fig:46yr_era5_training_metrics_1deg_gmroff} we provide the training dynamics of the model without GMR. The same plot with GMR is shown in the main body of the text (see \autoref{fig:46yr_era5_training_metrics_1deg_gmron}).

\begin{figure}[H]
  \centering
  \includegraphics[width=0.45\textwidth]{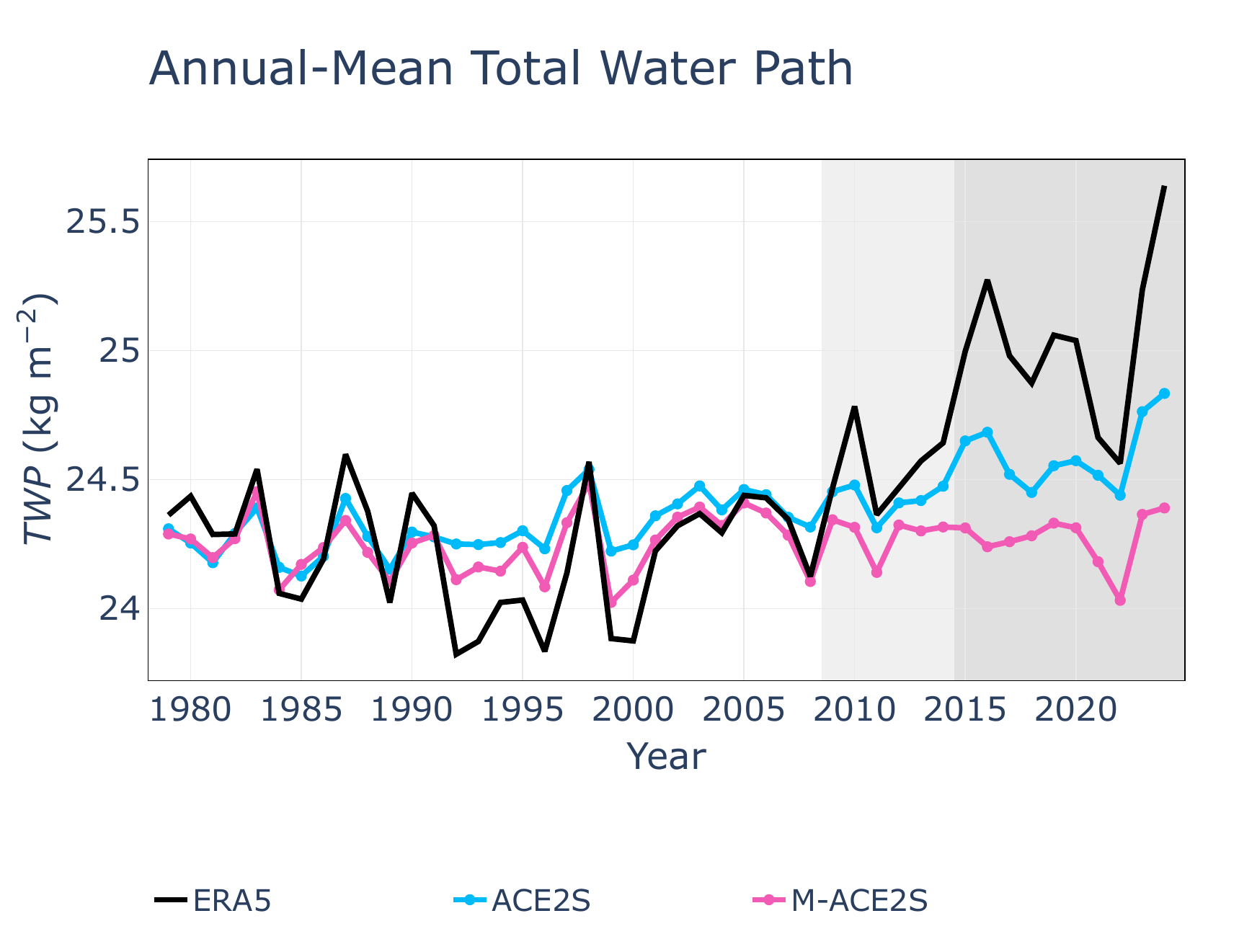}
  \includegraphics[width=0.45\textwidth]{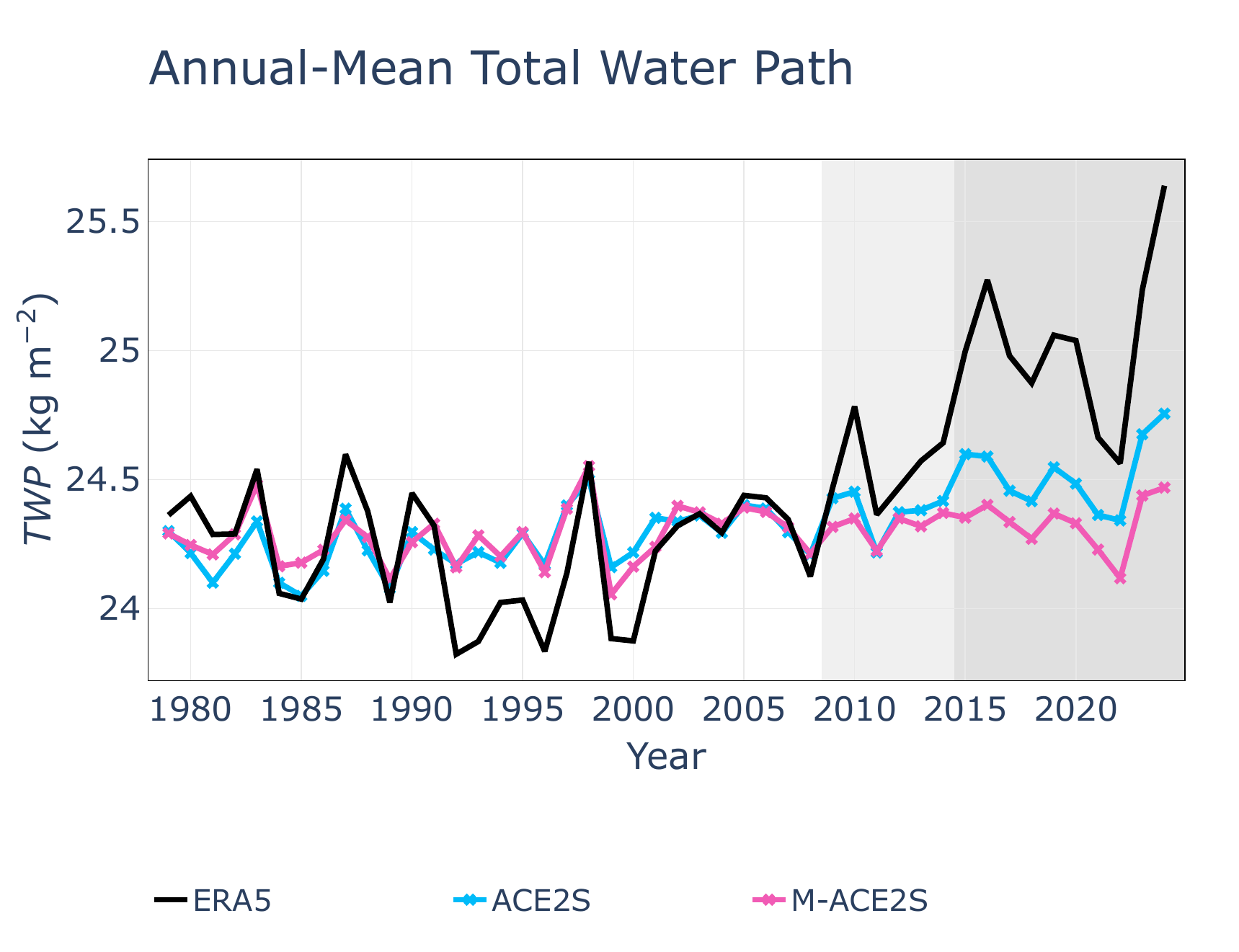}
  \caption{Global mean timeseries of total water path $TWP$ for ERA5 (black), M-ACE2S (pink), and ACE2S (blue). Shown are models trained without global mean removal (GMR) (left) and with global mean removal (GMR) (right). The plotted line is the mean of the eight-member initial-condition ensemble for each model. Results are for 1\textdegree\ 6-hourly models; see \autoref{fig:46yr_era5_TWP_4deg} for the 4\textdegree\ counterpart. Light and dark shading mark the AIMIP validation (2009--2014) and test (2015--2024) periods.}
  \label{fig:46yr_era5_TWP_1deg}
\end{figure}

In \autoref{fig:46yr_era5_TWP_1deg} we show the global mean timeseries of total water path $TWP$, for masked and unmasked models trained both with and without global mean removal. The corresponding 4\textdegree\ results are given in \autoref{appendix:additional_4deg_results}.

\if0
\begin{figure}[H]
  \centering
  \includegraphics[width=0.45\textwidth]{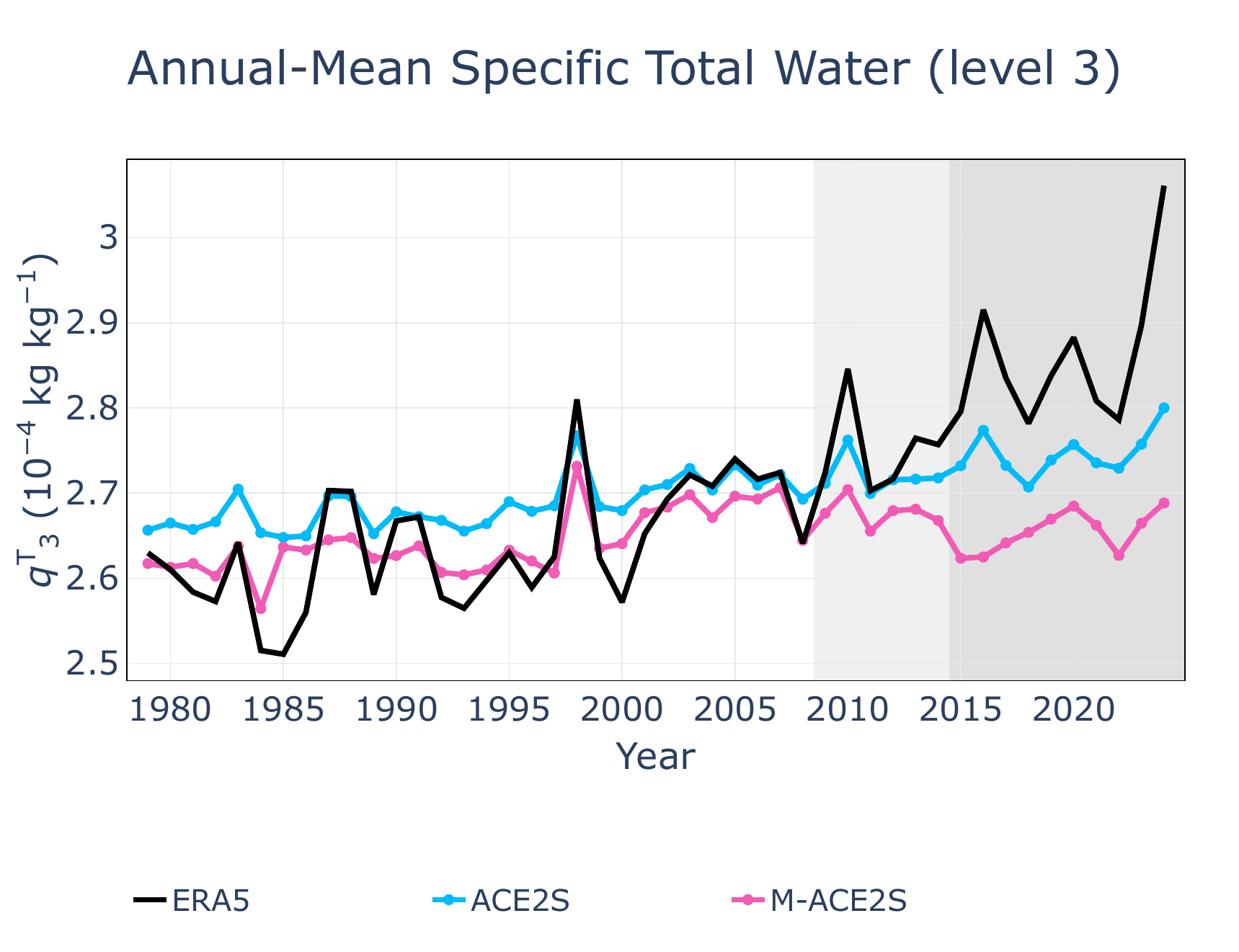}
  \includegraphics[width=0.45\textwidth]{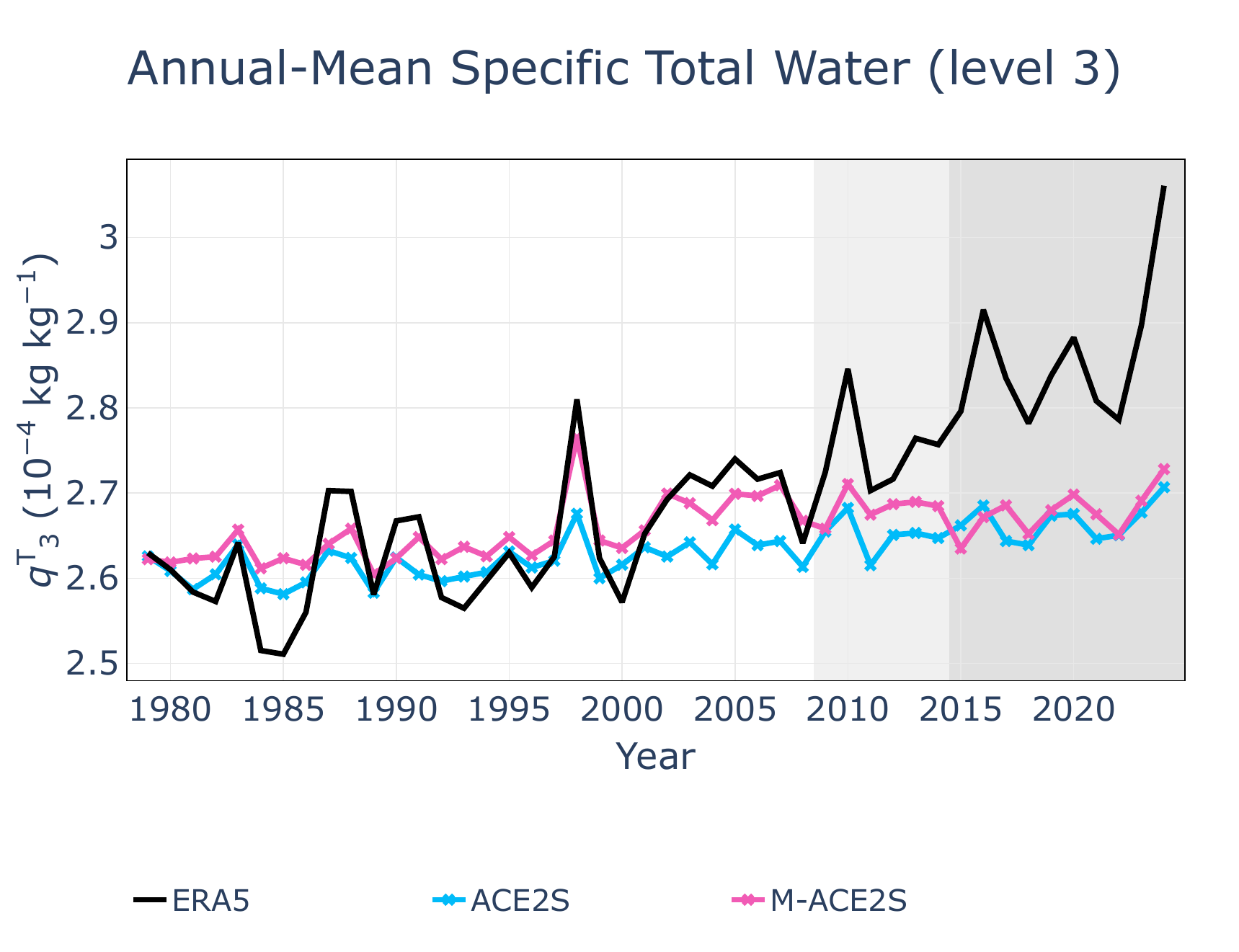}
  \caption{Global mean timeseries of upper troposphere specific total water $q^T_3$ for ERA5 (black), M-ACE2S (pink), and ACE2S (blue). Shown are models trained without global mean removal (GMR) (left) and with global mean removal (GMR) (right). The plotted line is the mean of the eight-member initial-condition ensemble for each model. Results are for 1\textdegree\ 6-hourly models; see \autoref{fig:46yr_era5_Q3_4deg} for the 4\textdegree\ counterpart.}
  \label{fig:46yr_era5_Q3_1deg}
\end{figure}
\fi

\autoref{table:twp_trend_1deg} scores the $TWP$ series of this figure as in \autoref{table:t7_trend_1deg}, for masked and unmasked models trained both with and without global mean removal. Masking lowers the training-period RMSE and trend error but raises every test-period error, with and without GMR; both models drift dry over the test period.

\begin{table}[H]
  \caption{Errors of the annual global-mean $TWP$ of the 46-year ERA5 rollouts of the 1\textdegree{} models over the AIMIP training (1979--2008) and test (2015--2024) periods. RMSE: root-mean-square of model minus ERA5 over the years of the period. Bias: mean of model minus ERA5. Trend error: model minus ERA5 least-squares linear trend. Each model is a single training run; bold marks the smaller magnitude of each ACE2S / M-ACE2S pair.}
  \label{table:twp_trend_1deg}
  \centering
  \small
  \begin{tabular}{lcccccc}
    \toprule
     & \multicolumn{2}{c}{RMSE (kg\,m$^{-2}$)} & \multicolumn{2}{c}{Bias (kg\,m$^{-2}$)} & \multicolumn{2}{c}{Trend error (kg\,m$^{-2}$\,decade$^{-1}$)} \\
    \cmidrule(lr){2-3} \cmidrule(lr){4-5} \cmidrule(lr){6-7}
     & Train & Test & Train & Test & Train & Test \\
    \midrule
    ACE2S (GMR off) & 0.20 & \textbf{0.47} & 0.09 & \textbf{$-$0.43} & 0.085 & \textbf{$-$0.045} \\
    M-ACE2S (GMR off) & \textbf{0.14} & 0.79 & \textbf{0.03} & $-$0.76 & \textbf{0.039} & $-$0.160 \\
    \midrule
    ACE2S (GMR on) & 0.18 & \textbf{0.54} & \textbf{0.04} & \textbf{$-$0.51} & 0.077 & \textbf{$-$0.094} \\
    M-ACE2S (GMR on) & \textbf{0.17} & 0.73 & 0.06 & $-$0.70 & \textbf{0.045} & $-$0.165 \\
    \bottomrule
  \end{tabular}
\end{table}

\if0
\begin{figure}[H]
  \centering
  \includegraphics[width=0.45\textwidth]{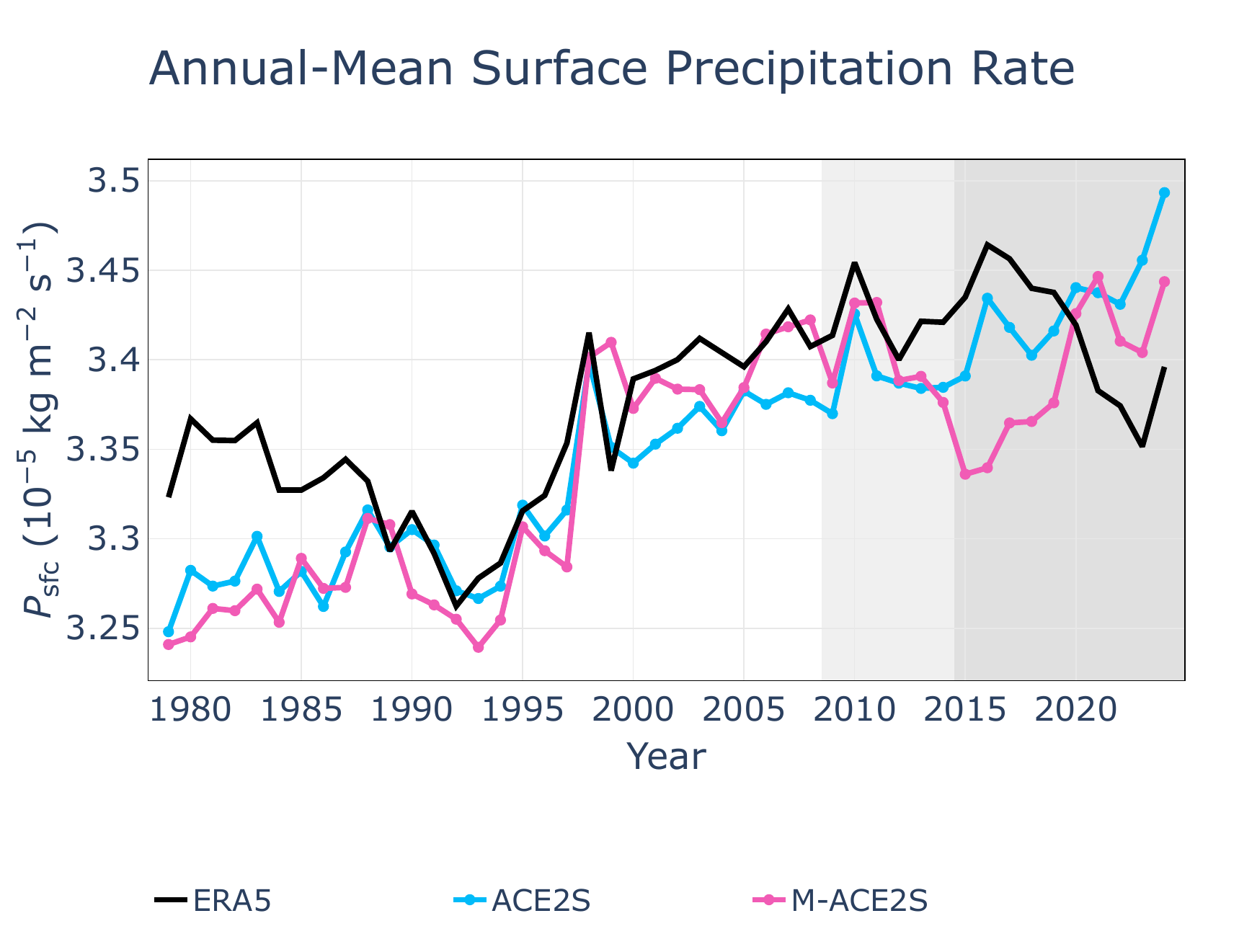}
  \includegraphics[width=0.45\textwidth]{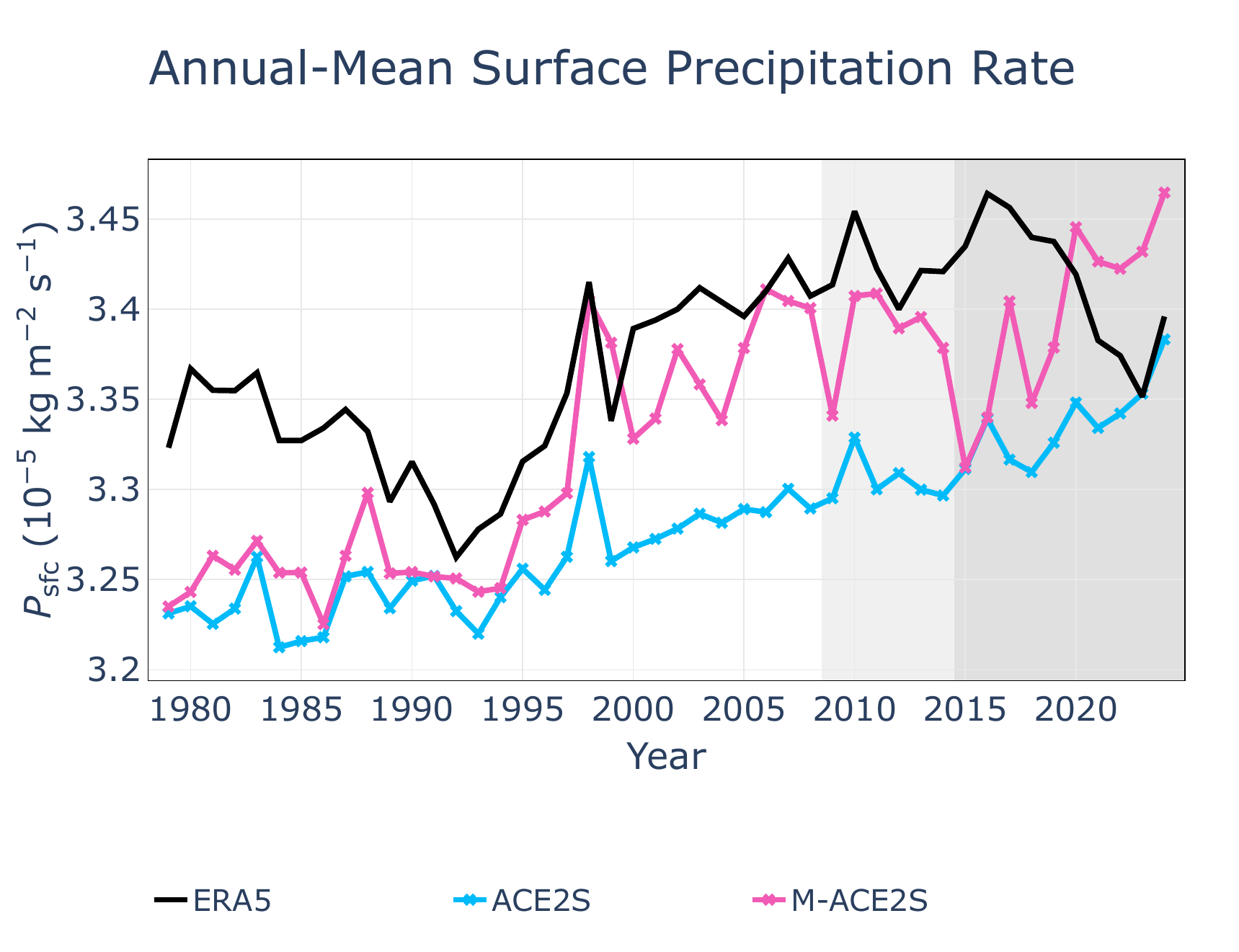}
  \caption{Global mean timeseries of surface precipitation $Psfc$ for ERA5 (black), M-ACE2S (pink), and ACE2S (blue). Shown are models trained without global mean removal (GMR) (left) and with global mean removal (GMR) (right). The plotted line is the mean of the eight-member initial-condition ensemble for each model. Results are for 1\textdegree\ 6-hourly models; see \autoref{fig:46yr_era5_Psfc_4deg} for the 4\textdegree\ counterpart.}
  \label{fig:46yr_era5_Psfc_1deg}
\end{figure}
\fi

\subsection{Time Mean Variable Plots}
\label{appendix:time_mean_variable_plots_1deg}

\begin{figure}[H]
  \centering
  \includegraphics[width=0.95\textwidth]{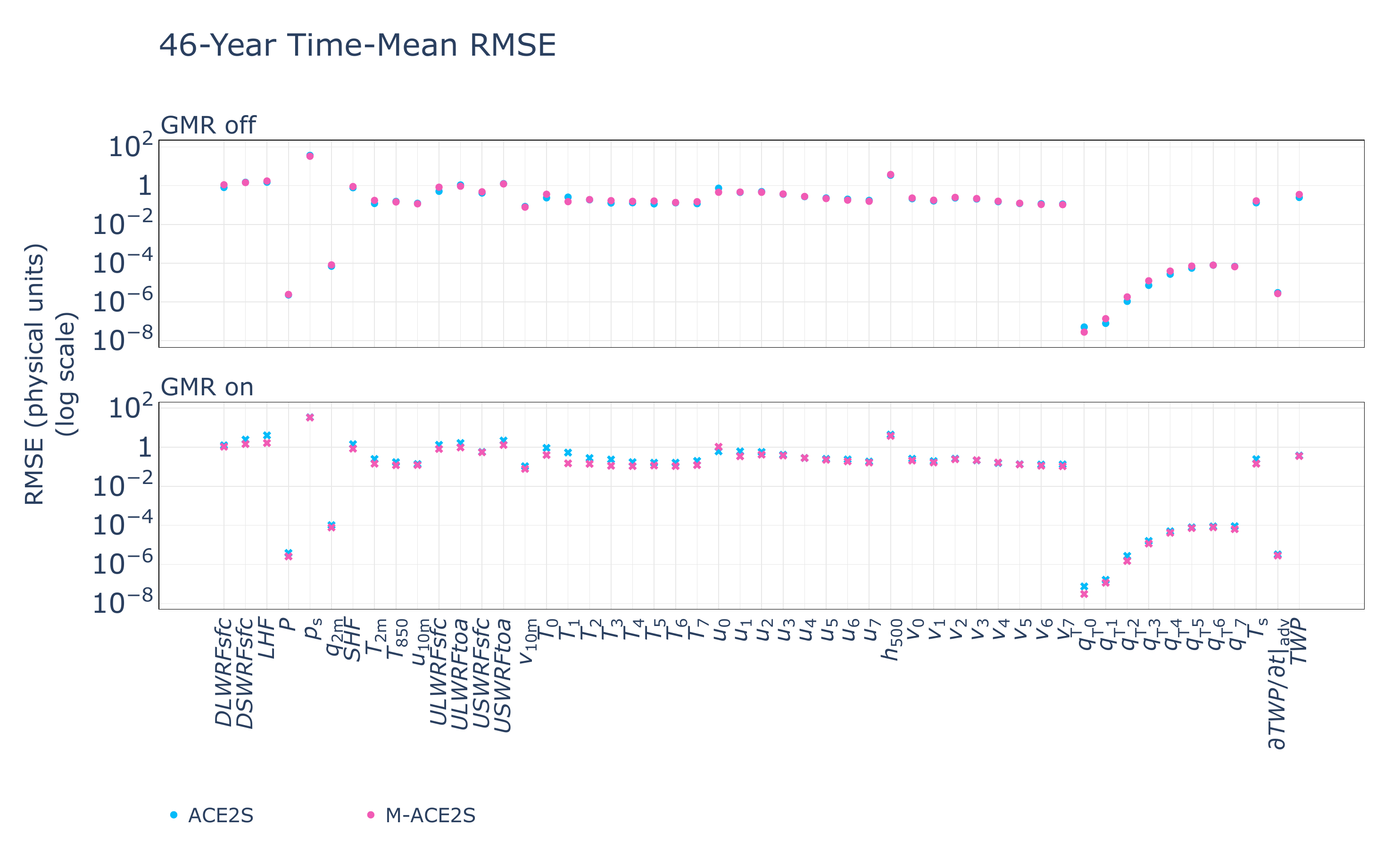}
  \caption{The time-mean RMSE of the model's 46 year rollout on ERA5 for all output variables without (top) and with (bottom) global mean removal. Results are for 1\textdegree\ 6-hourly models; see \autoref{fig:46yr_time_mean_rmse_era5_4deg} for the 4\textdegree\ counterpart.}
  \label{fig:46yr_time_mean_rmse_era5_1deg}
\end{figure}

Here we provide a plot of the time-mean RMSE for each variable for the 46 year forecast. The results are shown in \autoref{fig:46yr_time_mean_rmse_era5_1deg}. The 4\textdegree\ counterpart is \autoref{fig:46yr_time_mean_rmse_era5_4deg}. With global mean removal, masking lowers the time-mean RMSE for most fields (47 of 51); without it, masking lowers the RMSE for only 18 of 51 fields.

\section{Additional 4\textdegree\ Results}
\label{appendix:additional_4deg_results}

Here we evaluate 46-year climate rollouts on ERA5 for models trained on 4\textdegree\ daily data. Since climate forecasting is a nonlinear chaotic system \cite{Lorenz1963deterministic, slingo2011uncertainty}, a single forecast is not sufficient to represent the statistics of the system. For this reason, we trained 5 different random seeds from scratch for both M-ACE2S and ACE2S. The 4\textdegree\ daily models are cheaper to train, which is what makes this seed ensemble affordable.

These models differ from the 1\textdegree\ models of the main text in four ways, and in no others. First, the training data is a 4\textdegree\ ($45 \times 90$) daily-averaged split of ERA5 rather than a 1\textdegree\ ($180 \times 360$) 6-hourly one, so the model time step is $\Delta t = 1\,\mathrm{day}$. Second, the smaller grid shrinks the two resolution-dependent parameter blocks: the per-mode spectral weights and the learned positional embedding. These two changes cause the network to have about $2.0 \times 10^8$ trainable parameters rather than $8.0 \times 10^8$. Every architecture hyperparameter of \autoref{table:sfnoparams} is unchanged. Third, there is no multi-step fine-tuning stage: training is the 150 single-step epochs of \autoref{table:optimizationparams} only, spread over 2 GPUs rather than 8, and the reported model is the checkpoint selected over those epochs (\autoref{appendix:checkpoint_selection}). Fourth, each 4\textdegree\ configuration is trained from scratch with 5 different random seeds, so mean and standard deviation across seeds are reported, whereas the 1\textdegree\ results use one seed per configuration.

\begin{figure}[H]
  \centering
  \includegraphics[width=0.45\textwidth]{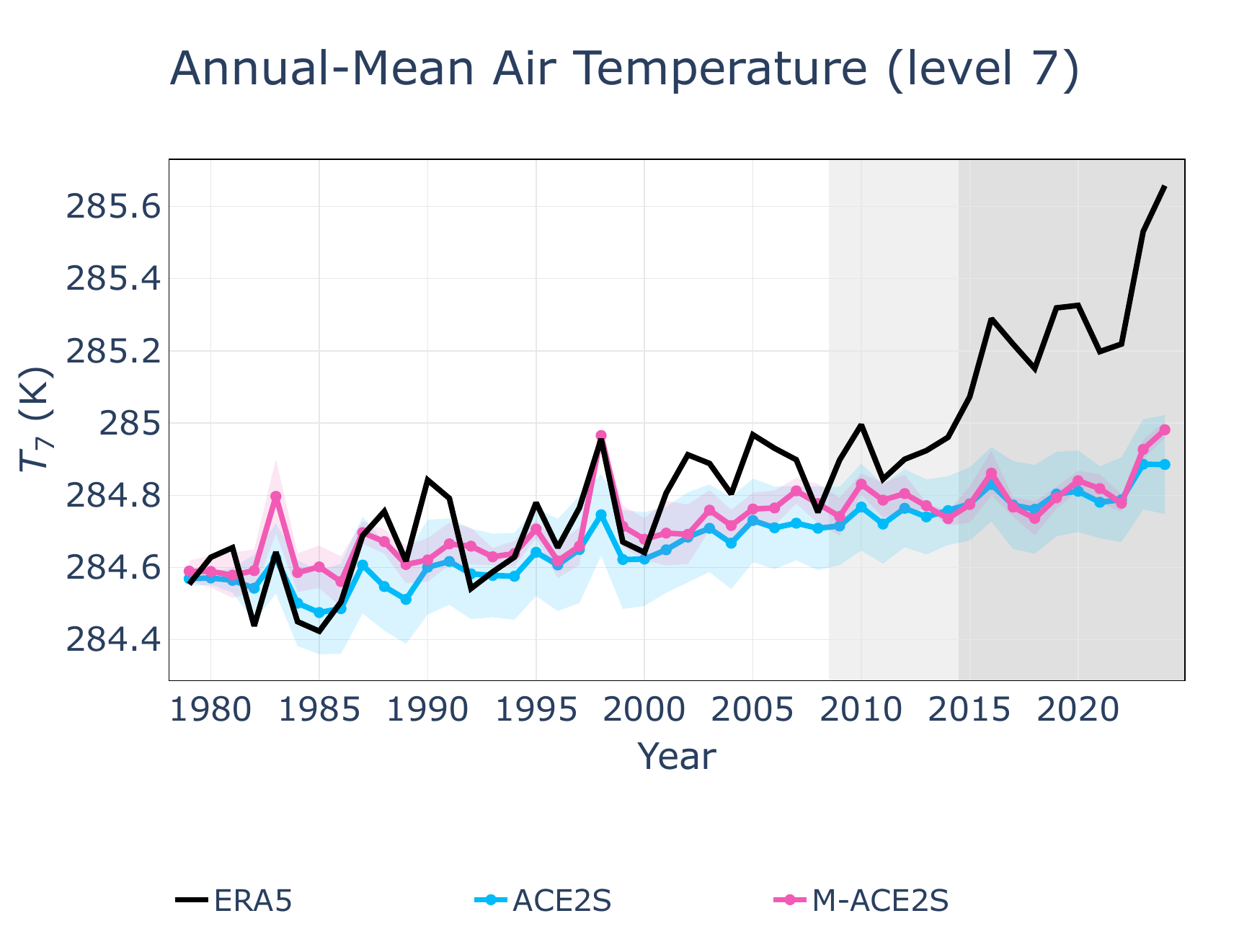}
  \includegraphics[width=0.45\textwidth]{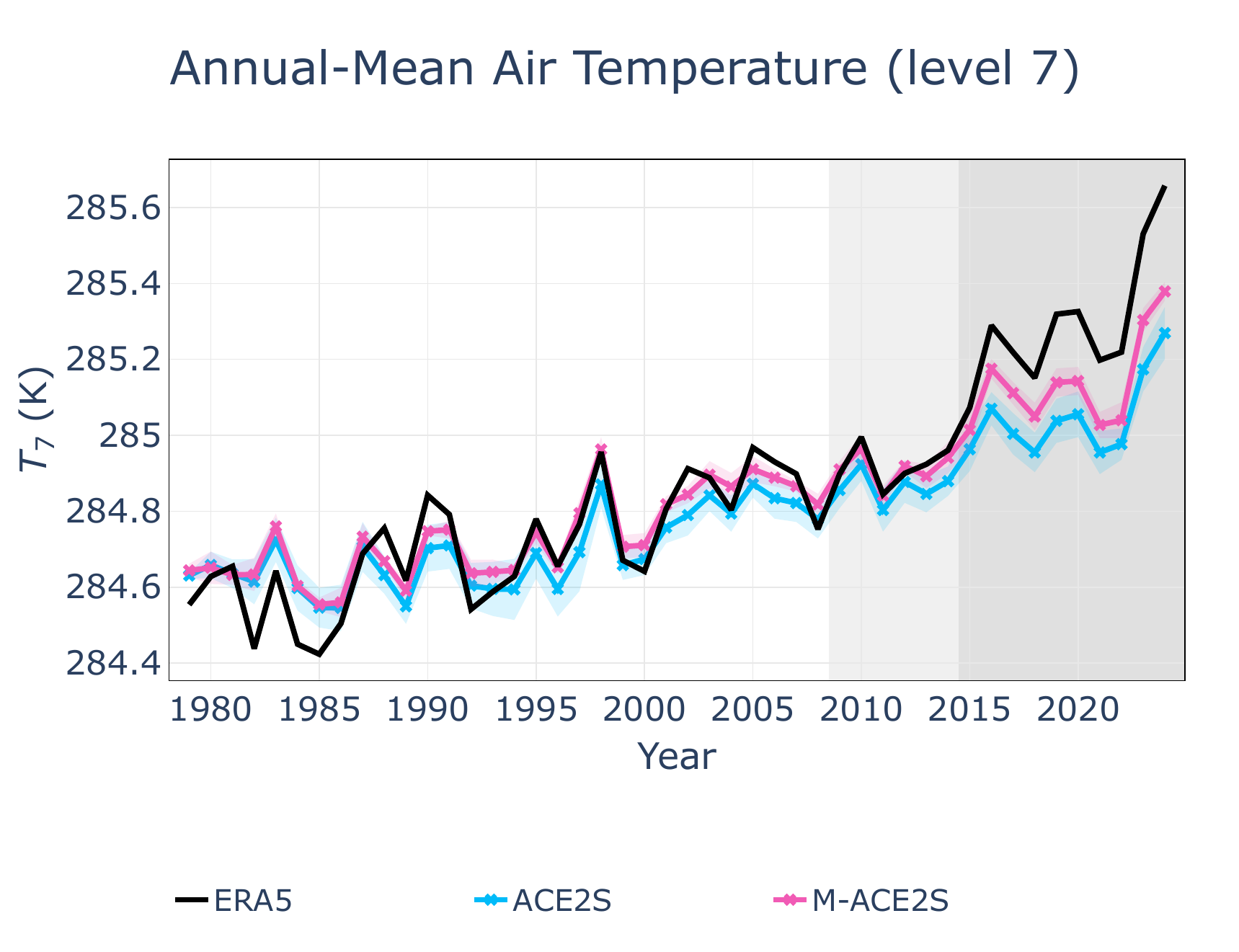}
  \caption{Global mean timeseries of near-surface air temperature $T_7$ for ERA5 (black), M-ACE2S (pink), and ACE2S (blue). Shown are models trained without global mean removal (GMR) (left) and with global mean removal (GMR) (right). Mean and standard deviation are computed over the ensemble of 5 random seed runs. Results are for 4\textdegree\ daily models; see \autoref{fig:46yr_era5_T7_gmroff_1deg} and \autoref{fig:46yr_era5_T7_1deg} for the 1\textdegree\ counterparts without and with GMR, respectively. Light and dark shading mark the AIMIP validation (2009--2014) and test (2015--2024) periods.}
  \label{fig:46yr_era5_T7_4deg}
\end{figure}

\autoref{fig:46yr_era5_T7_4deg} shows long-time forecasts of $T_7$ on ERA5 for M-ACE2S and ACE2S trained with and without global mean removal. We observe that, without global mean removal, masking acts as a regularizer, reducing the spread of the 5-seed ensemble for the $T_7$ forecast across 46 years. As a result of training M-ACE2S with our simple masking strategy, the model is able to better follow the long-term climatology of the near-surface air temperature $T_7$ compared to the unmasked 5-seed ensemble. For the models trained with global mean removal, both the masked and unmasked models are able to capture the long-term climatology of $T_7$ with improved accuracy, but the masked model performs slightly better in-sample and out-of-sample.

\autoref{table:t7_trend_4deg} repeats \autoref{table:t7_trend_1deg} and \autoref{table:t7_trend_1deg_gmroff} for the 4\textdegree\ models, averaged over five seeds. The ordering matches: M-ACE2S with GMR has the smallest test-period RMSE, bias, and trend error, and masking does not widen the seed-to-seed spread in any column except the test-period trend error without GMR.

\begin{table}[H]
  \caption{Errors of the annual global-mean $T_7$ of the 46-year ERA5 rollouts of the 4\textdegree{} models over the AIMIP training (1979--2008) and test (2015--2024) periods. RMSE: root-mean-square of model minus ERA5 over the years of the period. Bias: mean of model minus ERA5. Trend error: model minus ERA5 least-squares linear trend. Values are mean $\pm$ standard deviation over 5 seeds; bold marks the smaller magnitude of each ACE2S / M-ACE2S pair.}
  \label{table:t7_trend_4deg}
  \centering
  \small
  \resizebox{\linewidth}{!}{\begin{tabular}{lcccccc}
    \toprule
     & \multicolumn{2}{c}{RMSE (K)} & \multicolumn{2}{c}{Bias (K)} & \multicolumn{2}{c}{Trend error (K\,decade$^{-1}$)} \\
    \cmidrule(lr){2-3} \cmidrule(lr){4-5} \cmidrule(lr){6-7}
     & Train & Test & Train & Test & Train & Test \\
    \midrule
    ACE2S (GMR off) & 0.16 $\pm$ 0.05 & 0.51 $\pm$ 0.11 & $-$0.09 $\pm$ 0.11 & $-$0.49 $\pm$ 0.11 & \textbf{$-$0.060} $\pm$ 0.028 & $-$0.339 $\pm$ 0.044 \\
    M-ACE2S (GMR off) & \textbf{0.12} $\pm$ 0.03 & \textbf{0.48} $\pm$ 0.02 & \textbf{$-$0.03} $\pm$ 0.04 & \textbf{$-$0.47} $\pm$ 0.02 & $-$0.064 $\pm$ 0.009 & \textbf{$-$0.268} $\pm$ 0.086 \\
    \midrule
    ACE2S (GMR on) & 0.10 $\pm$ 0.01 & 0.26 $\pm$ 0.05 & \textbf{$-$0.02} $\pm$ 0.05 & $-$0.25 $\pm$ 0.05 & $-$0.050 $\pm$ 0.007 & $-$0.227 $\pm$ 0.044 \\
    M-ACE2S (GMR on) & \textbf{0.08} $\pm$ 0.01 & \textbf{0.17} $\pm$ 0.03 & 0.03 $\pm$ 0.02 & \textbf{$-$0.16} $\pm$ 0.03 & \textbf{$-$0.031} $\pm$ 0.004 & \textbf{$-$0.202} $\pm$ 0.040 \\
    \bottomrule
  \end{tabular}
  }
\end{table}

\begin{figure}[H]
  \centering
  \includegraphics[width=0.45\textwidth]{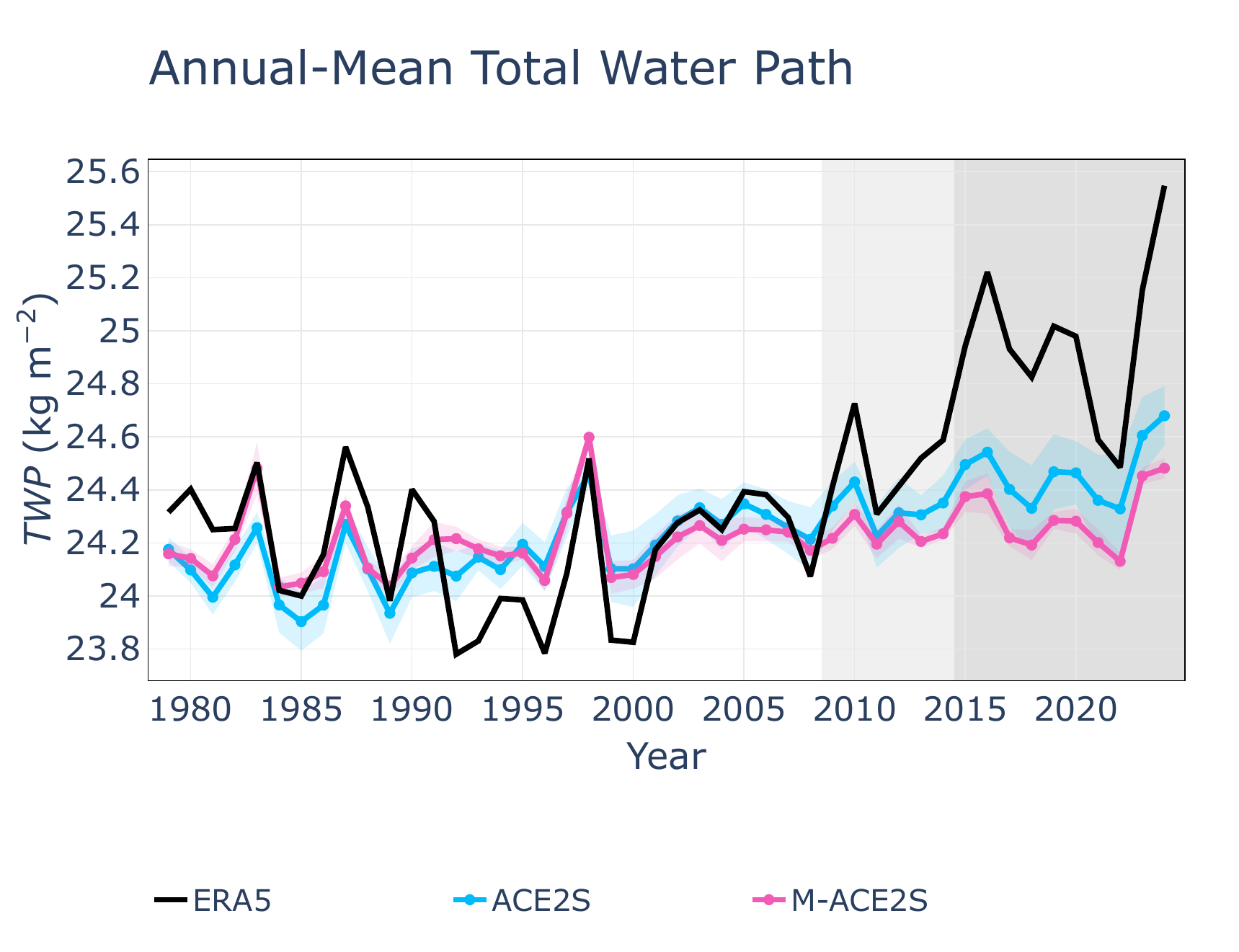}
  \includegraphics[width=0.45\textwidth]{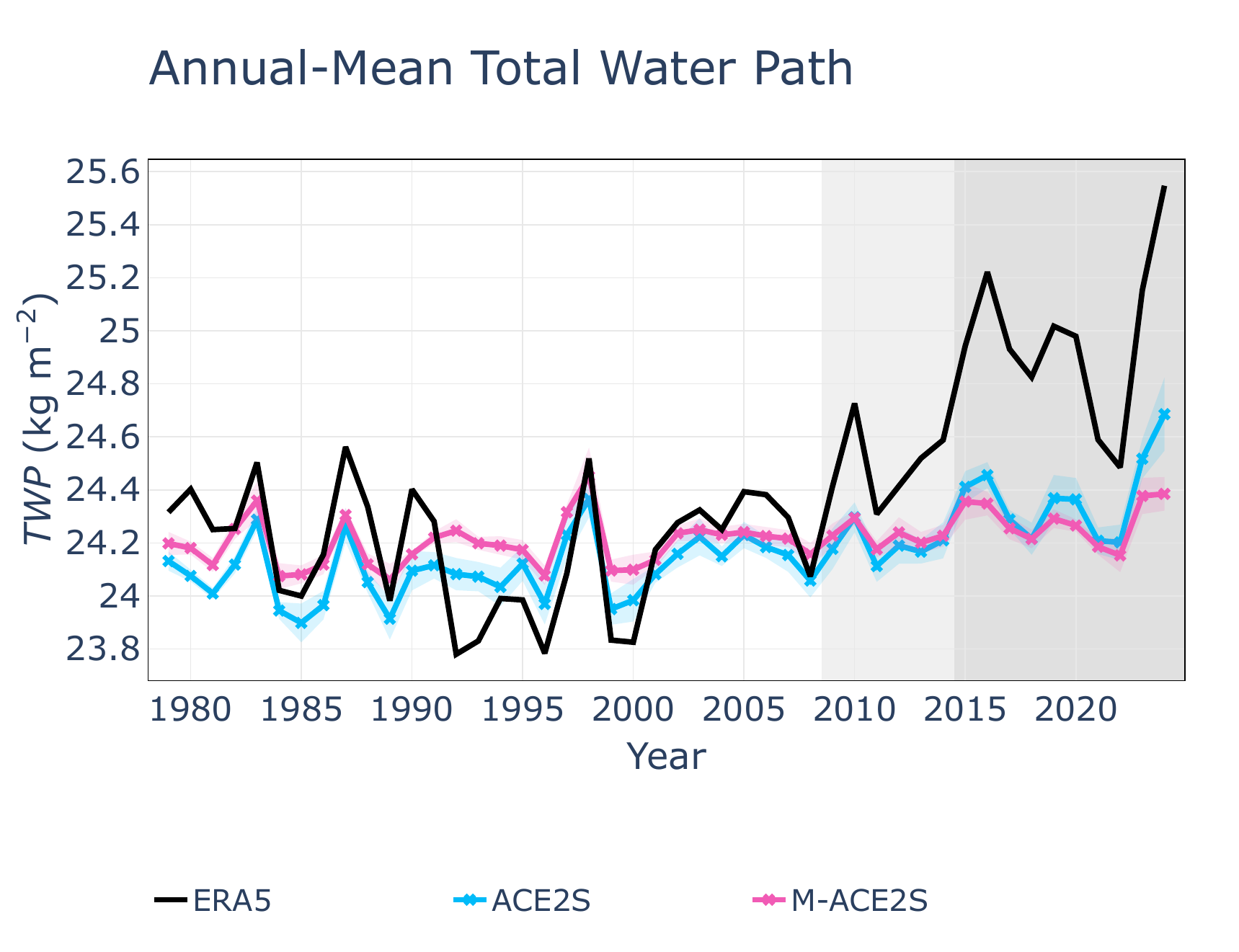}
  \caption{Global mean timeseries of total water path $TWP$ for ERA5 (black), M-ACE2S (pink), and ACE2S (blue). Shown are models trained without global mean removal (GMR) (left) and with global mean removal (GMR) (right). Mean and standard deviation are computed over the ensemble of 5 random seed runs. Results are for 4\textdegree\ daily models; see \autoref{fig:46yr_era5_TWP_1deg} for the 1\textdegree\ counterpart. Light and dark shading mark the AIMIP validation (2009--2014) and test (2015--2024) periods.}
  \label{fig:46yr_era5_TWP_4deg}
\end{figure}

In \autoref{fig:46yr_era5_TWP_4deg} we show the global mean timeseries of total water path $TWP$, for masked and unmasked models trained both with and without global mean removal.

\if0
\begin{figure}[H]
  \centering
  \includegraphics[width=0.45\textwidth]{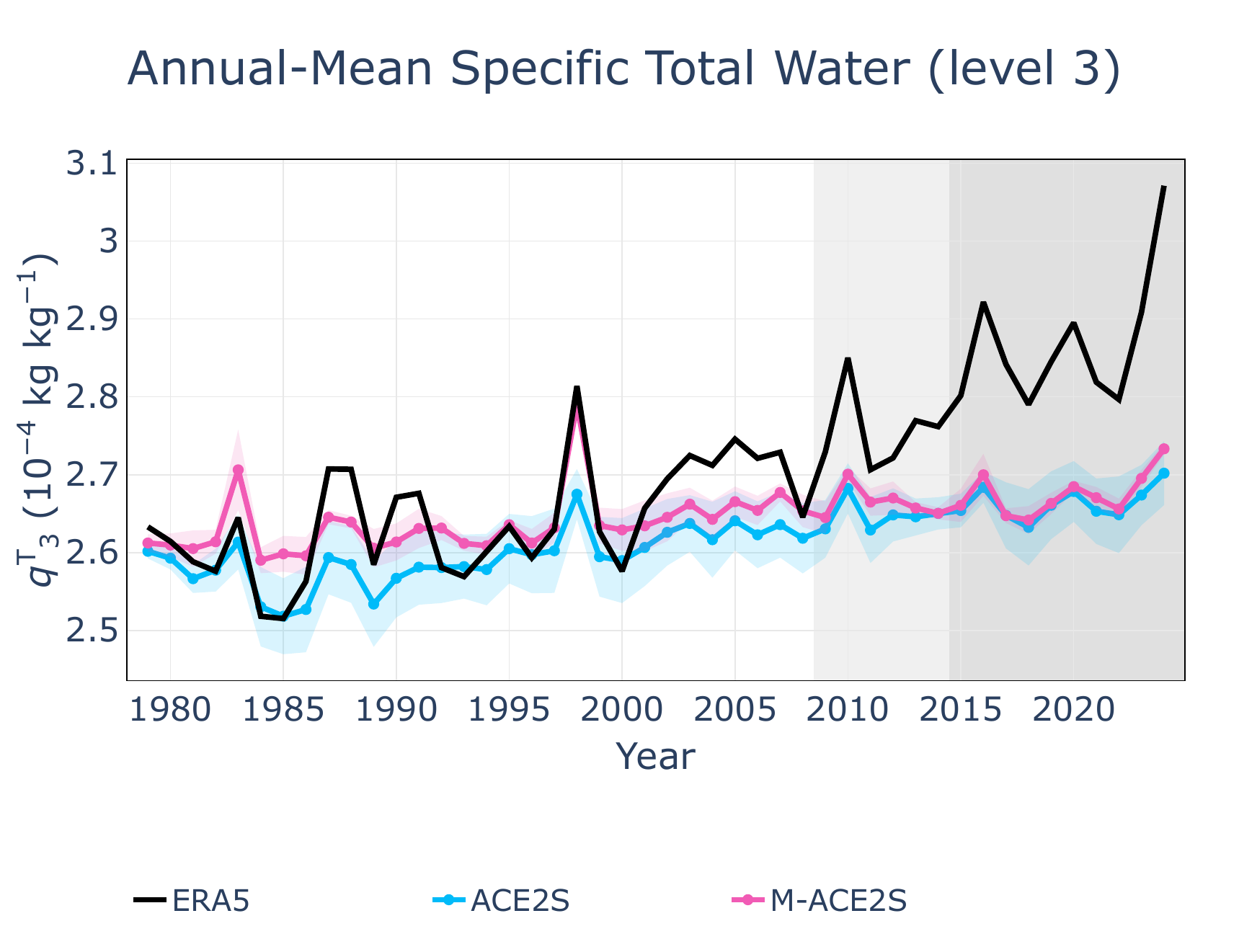}
  \includegraphics[width=0.45\textwidth]{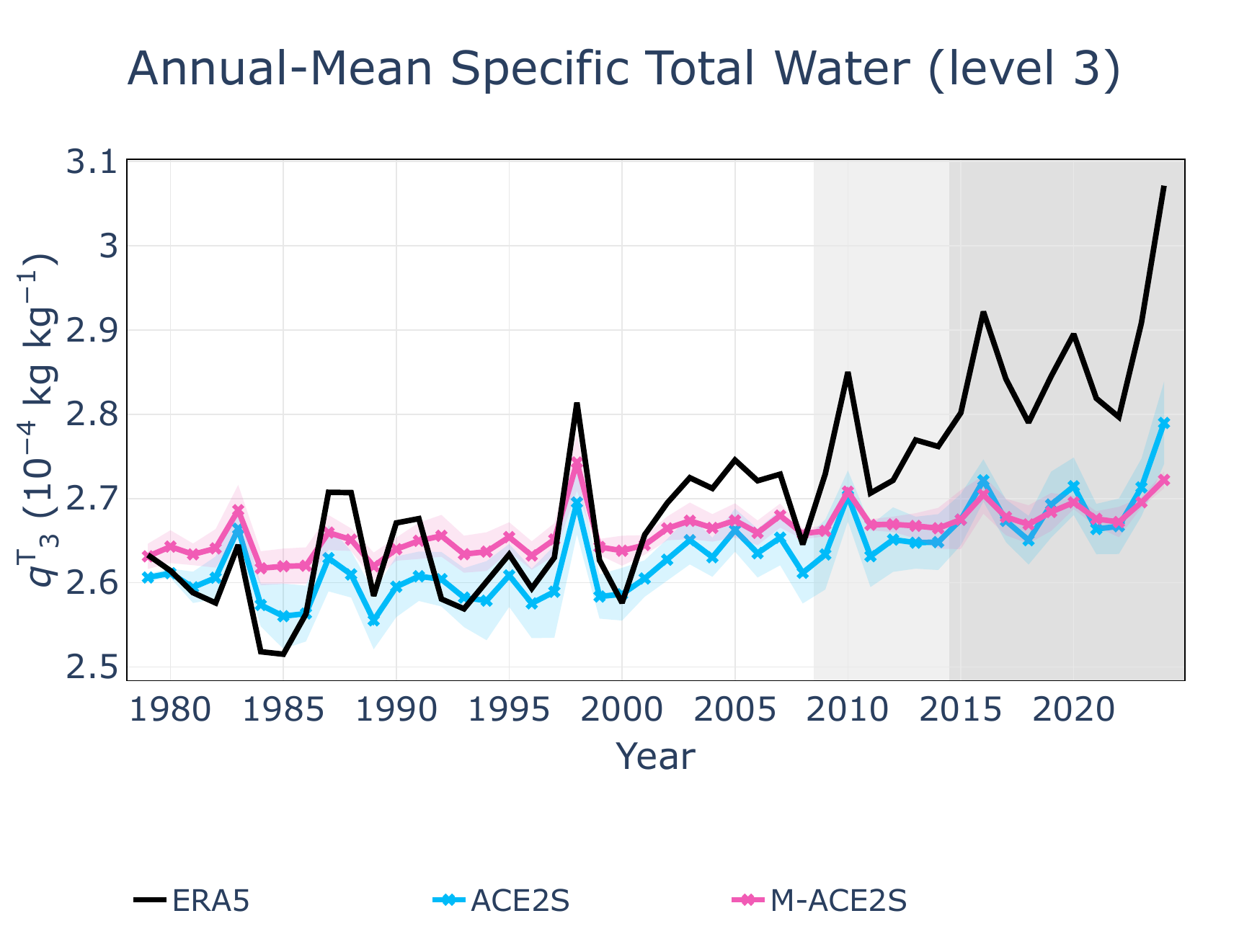}
  \caption{Global mean timeseries of upper troposphere specific total water $q^T_3$ for ERA5 (black), M-ACE2S (pink), and ACE2S (blue). Shown are models trained without global mean removal (GMR) (left) and with global mean removal (GMR) (right). Mean and standard deviation are computed over the ensemble of 5 random seed runs. Results are for 4\textdegree\ daily models; see \autoref{fig:46yr_era5_Q3_1deg} for the 1\textdegree\ counterpart.}
  \label{fig:46yr_era5_Q3_4deg}
\end{figure}
\fi

\autoref{table:twp_trend_4deg} scores the $TWP$ series of this figure as in \autoref{table:t7_trend_4deg}, averaged over five seeds. It matches the 1\textdegree\ result: masking lowers the training-period trend error but raises the test-period errors in both GMR settings.

\begin{table}[H]
  \caption{Errors of the annual global-mean $TWP$ of the 46-year ERA5 rollouts of the 4\textdegree{} models over the AIMIP training (1979--2008) and test (2015--2024) periods. RMSE: root-mean-square of model minus ERA5 over the years of the period. Bias: mean of model minus ERA5. Trend error: model minus ERA5 least-squares linear trend. Values are mean $\pm$ standard deviation over 5 seeds; bold marks the smaller magnitude of each ACE2S / M-ACE2S pair.}
  \label{table:twp_trend_4deg}
  \centering
  \small
  \resizebox{\linewidth}{!}{\begin{tabular}{lcccccc}
    \toprule
     & \multicolumn{2}{c}{RMSE (kg\,m$^{-2}$)} & \multicolumn{2}{c}{Bias (kg\,m$^{-2}$)} & \multicolumn{2}{c}{Trend error (kg\,m$^{-2}$\,decade$^{-1}$)} \\
    \cmidrule(lr){2-3} \cmidrule(lr){4-5} \cmidrule(lr){6-7}
     & Train & Test & Train & Test & Train & Test \\
    \midrule
    ACE2S (GMR off) & 0.21 $\pm$ 0.02 & \textbf{0.54} $\pm$ 0.12 & $-$0.02 $\pm$ 0.08 & \textbf{$-$0.50} $\pm$ 0.14 & 0.116 $\pm$ 0.033 & \textbf{$-$0.010} $\pm$ 0.057 \\
    M-ACE2S (GMR off) & \textbf{0.19} $\pm$ 0.01 & 0.70 $\pm$ 0.03 & \textbf{0.01} $\pm$ 0.04 & $-$0.67 $\pm$ 0.03 & \textbf{0.055} $\pm$ 0.012 & $-$0.059 $\pm$ 0.091 \\
    \midrule
    ACE2S (GMR on) & 0.19 $\pm$ 0.02 & \textbf{0.62} $\pm$ 0.06 & $-$0.08 $\pm$ 0.05 & \textbf{$-$0.60} $\pm$ 0.06 & 0.061 $\pm$ 0.012 & \textbf{0.028} $\pm$ 0.083 \\
    M-ACE2S (GMR on) & 0.19 $\pm$ 0.01 & 0.72 $\pm$ 0.02 & \textbf{0.02} $\pm$ 0.02 & $-$0.69 $\pm$ 0.02 & \textbf{0.042} $\pm$ 0.011 & $-$0.130 $\pm$ 0.099 \\
    \bottomrule
  \end{tabular}
  }
\end{table}

\if0
\begin{figure}[H]
  \centering
  \includegraphics[width=0.45\textwidth]{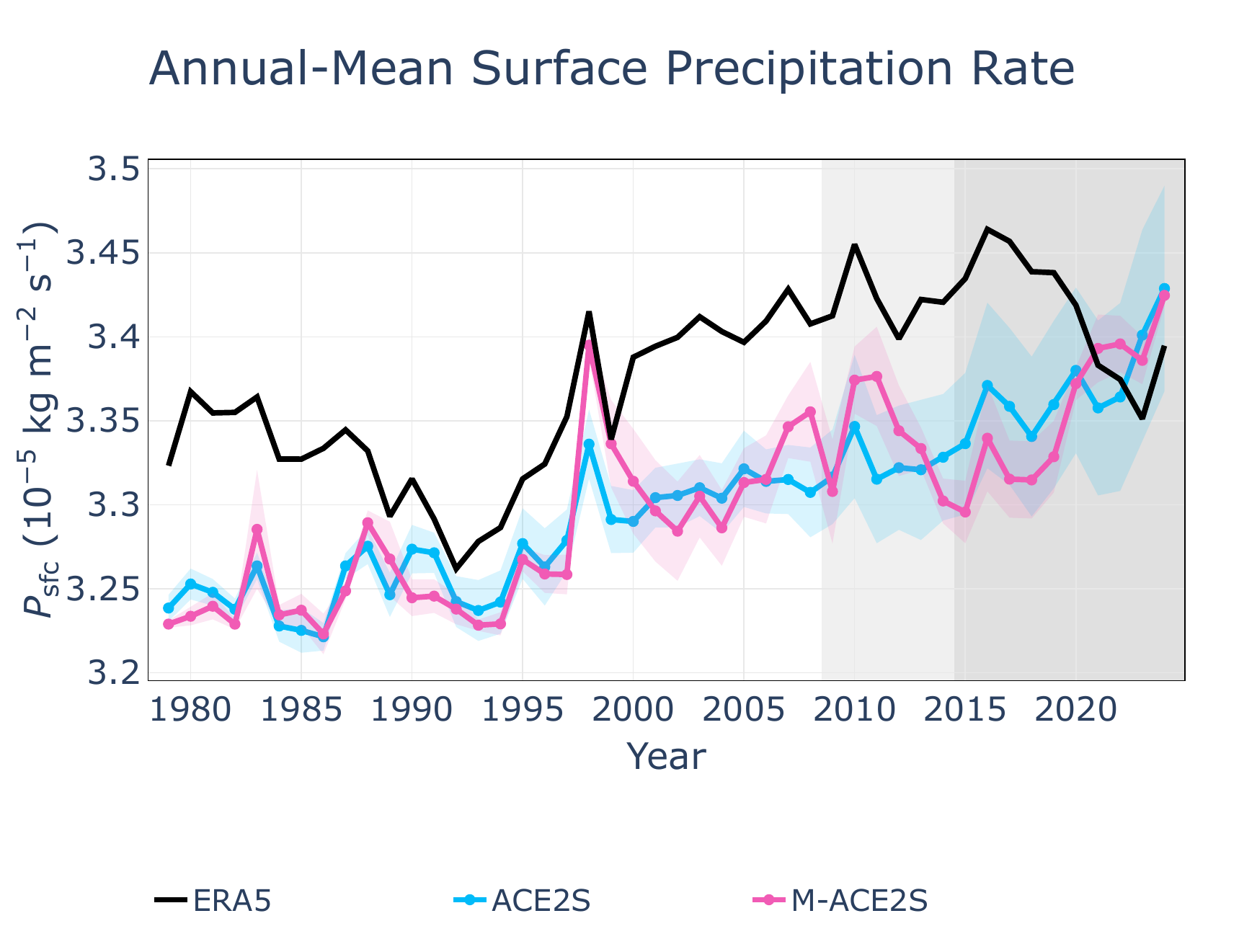}
  \includegraphics[width=0.45\textwidth]{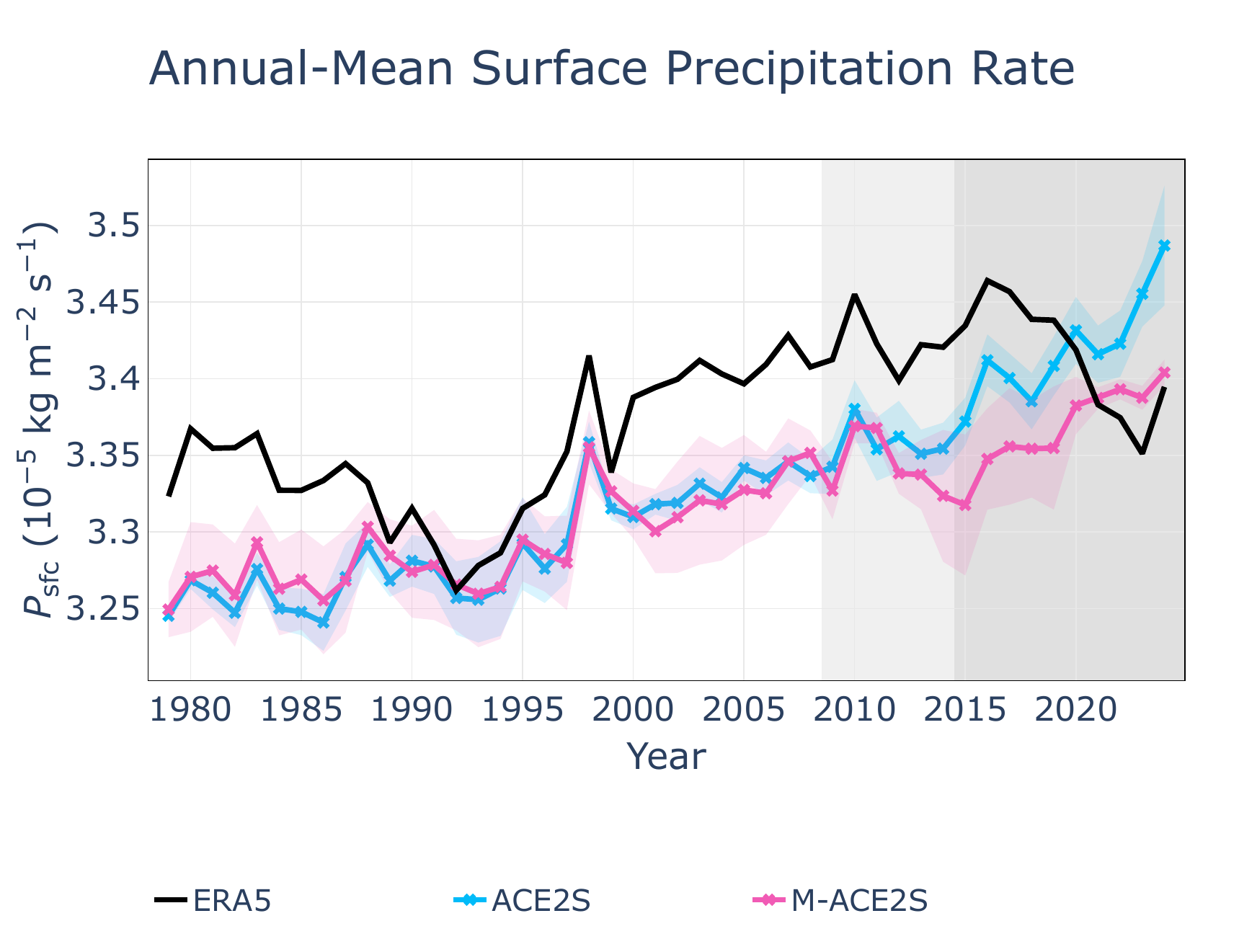}
  \caption{Global mean timeseries of surface precipitation $Psfc$ for ERA5 (black), M-ACE2S (pink), and ACE2S (blue). Shown are models trained without global mean removal (GMR) (left) and with global mean removal (GMR) (right). Mean and standard deviation are computed over the ensemble of 5 random seed runs. Results are for 4\textdegree\ daily models; see \autoref{fig:46yr_era5_Psfc_1deg} for the 1\textdegree\ counterpart.}
  \label{fig:46yr_era5_Psfc_4deg}
\end{figure}
\fi

\subsection{Time Mean Variable Plots}
\label{appendix:time_mean_variable_plots_4deg}

\begin{figure}[H]
  \centering
  \includegraphics[width=0.95\textwidth]{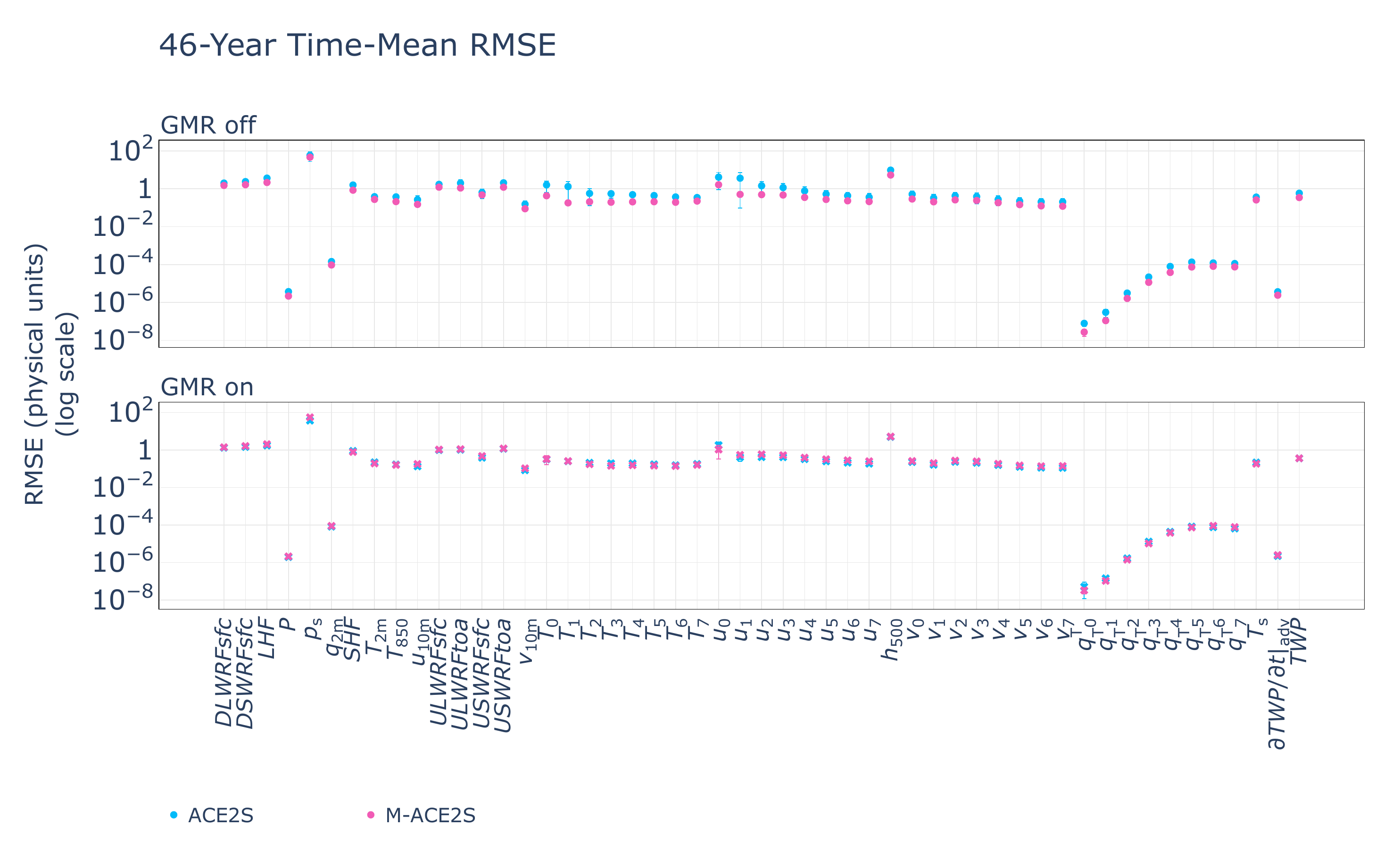}
  \caption{The time-mean RMSE of the model's 46 year rollout on ERA5 for all output variables without (top) and with (bottom) global mean removal. Results are for 4\textdegree\ daily models; see \autoref{fig:46yr_time_mean_rmse_era5_1deg} for the 1\textdegree\ counterpart.}
  \label{fig:46yr_time_mean_rmse_era5_4deg}
\end{figure}

Here we provide the 4\textdegree\ counterparts of the time-mean plots in \autoref{appendix:time_mean_variable_plots_1deg}: the time-mean RMSE for each variable over the 46 year forecast in \autoref{fig:46yr_time_mean_rmse_era5_4deg}. At 4\textdegree\ the split is the reverse of the 1\textdegree\ case: masking lowers the time-mean RMSE for every field without global mean removal (51 of 51), but for only 17 of 51 fields \textit{with} GMR. The per-field RMSE gains and the lower 5-year inference error of the main text appear to be specific to the 1\textdegree\ models (see also \autoref{appendix:training_metrics_4deg}), although M-ACE2S with GMR still has the smallest test-period $T_7$ RMSE, bias, and trend error across five seeds (\autoref{table:t7_trend_4deg}).

\if0
\begin{figure}[H]
  \centering
  \includegraphics[width=0.95\textwidth]{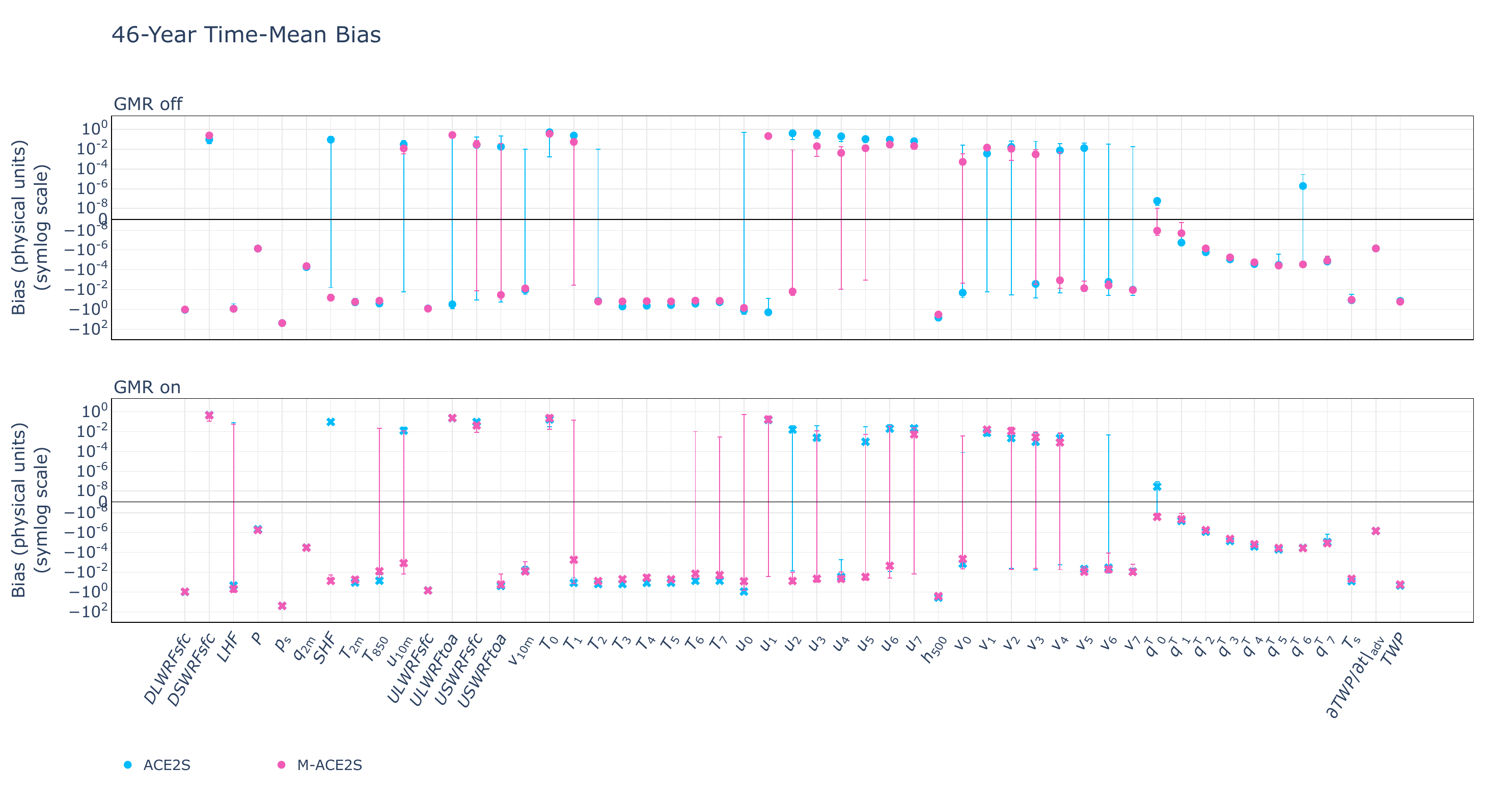}
  \caption{The time-mean bias of the model's 46 year rollout on ERA5 for all output variables without (top) and with (bottom) global mean removal. Results are for 4\textdegree\ daily models; see \autoref{fig:46yr_time_mean_bias_era5_1deg} for the 1\textdegree\ counterpart.}
  \label{fig:46yr_time_mean_bias_era5_4deg}
\end{figure}
\fi

\subsection{Training and Checkpoint Selection Metrics}
\label{appendix:training_metrics_4deg}

\begin{figure}[H]
  \centering
  \includegraphics[width=0.45\textwidth]{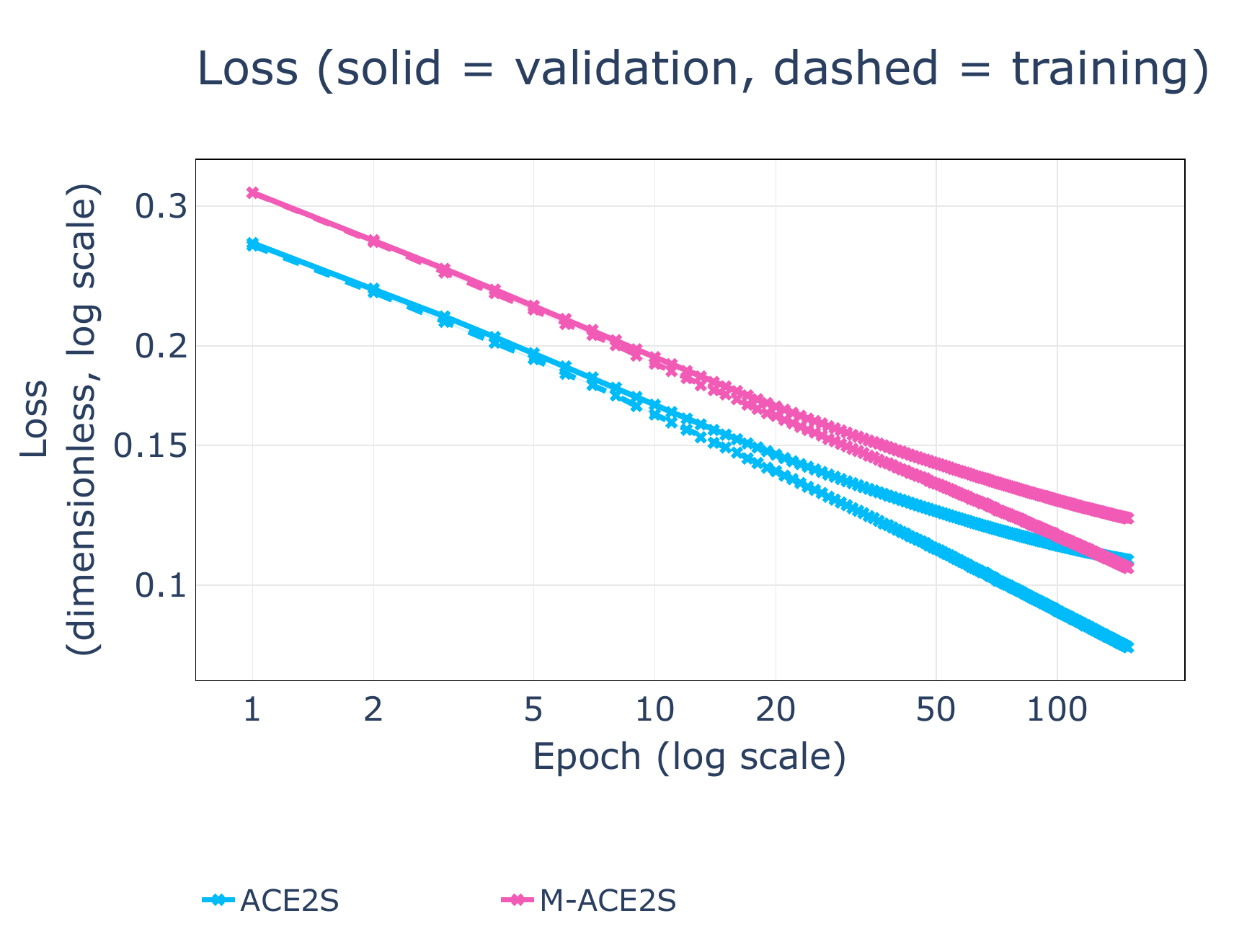}
  \includegraphics[width=0.45\textwidth]{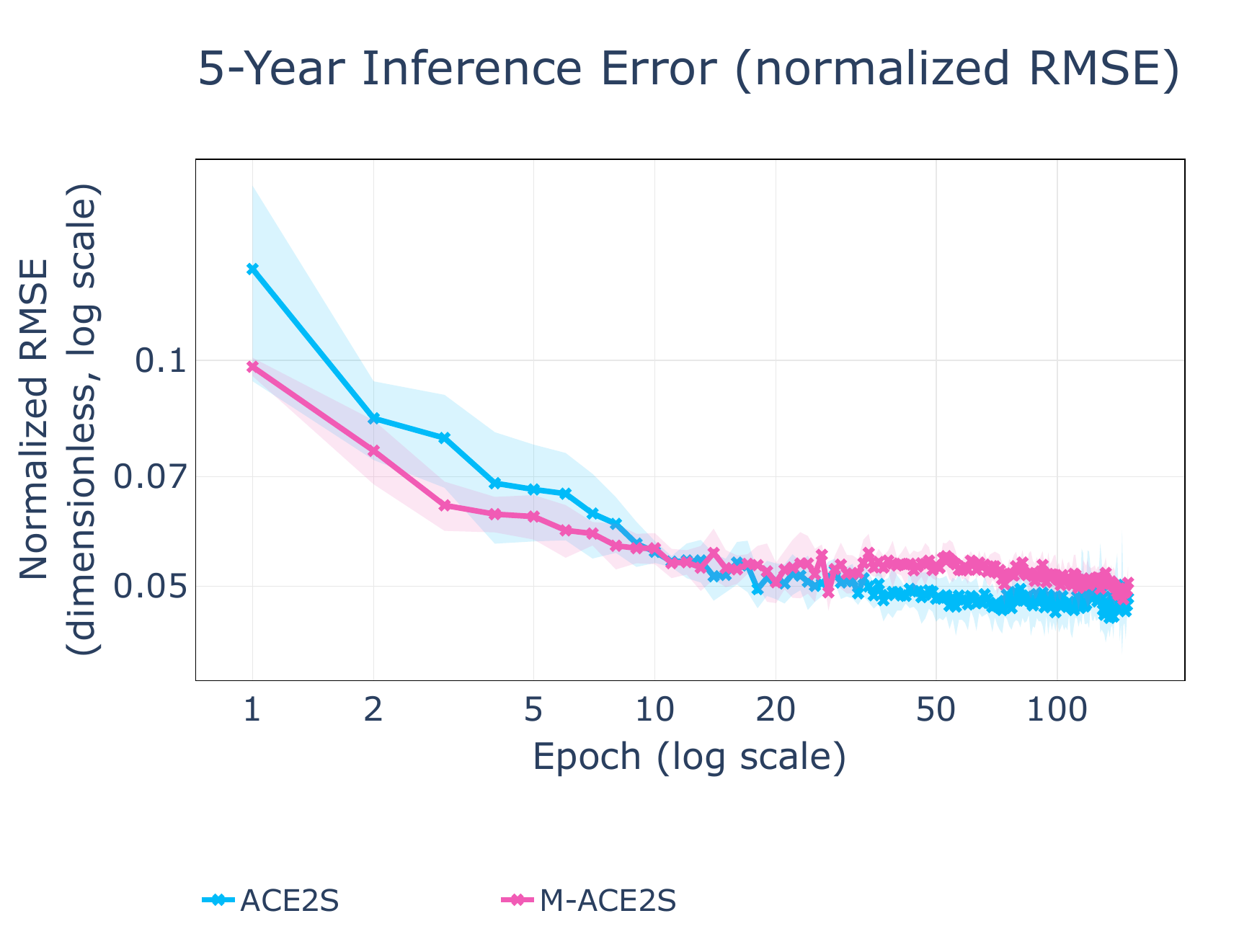}
  \caption{Training (dashed) and validation (solid) loss (left) and AIMIP checkpoint time-mean normalized RMSE averaged over channels (right) during training for M-ACE2S (pink) and ACE2S (blue) trained \emph{with} global mean removal (GMR on). Mean and standard deviation are computed over the ensemble of 5 random seed runs. Results are for 4\textdegree\ daily models; see \autoref{fig:46yr_era5_training_metrics_1deg_gmron} for the 1\textdegree\ counterpart.}
  \label{fig:46yr_era5_training_metrics_4deg_gmron}
\end{figure}

\begin{figure}[H]
  \centering
  \includegraphics[width=0.45\textwidth]{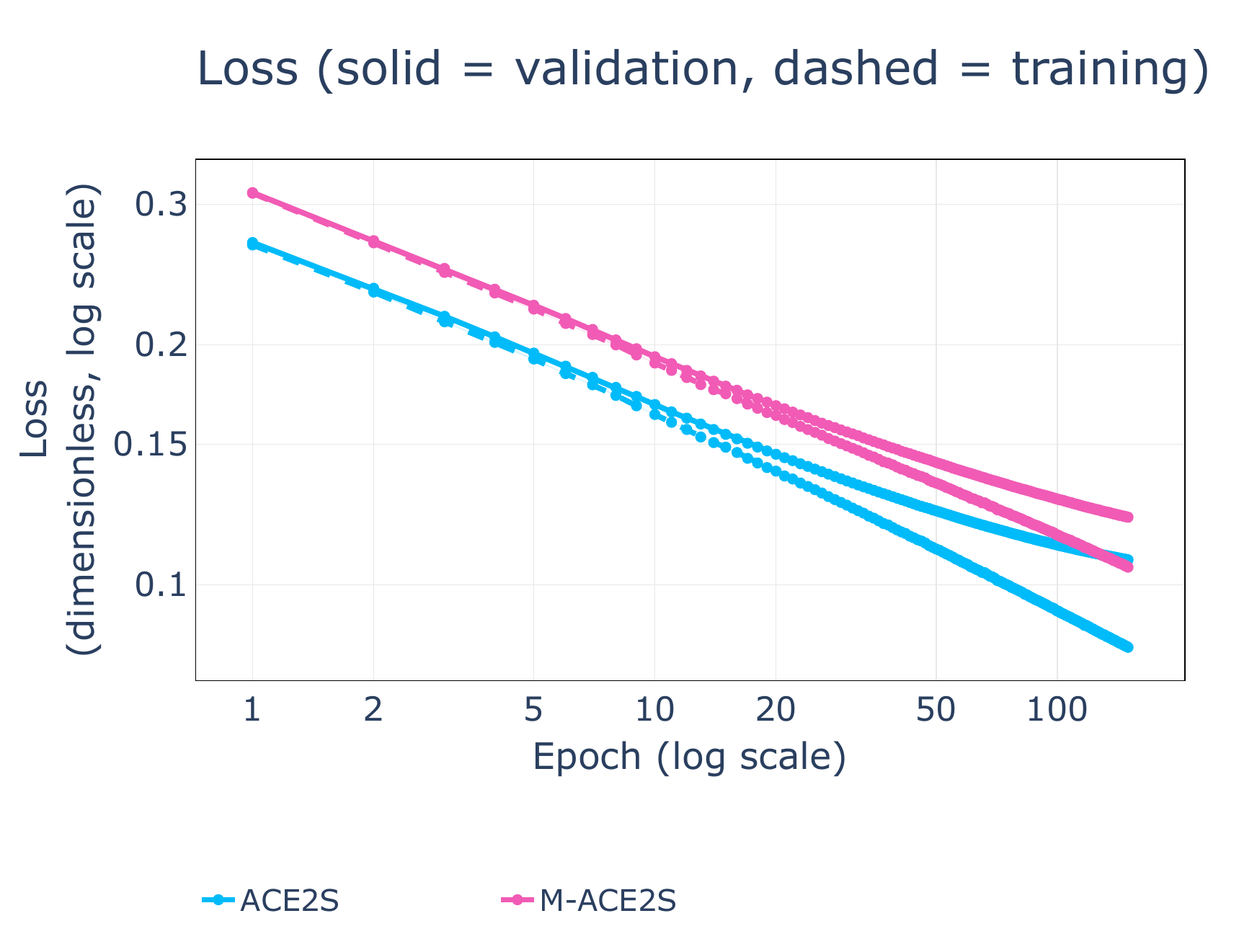}
  \includegraphics[width=0.45\textwidth]{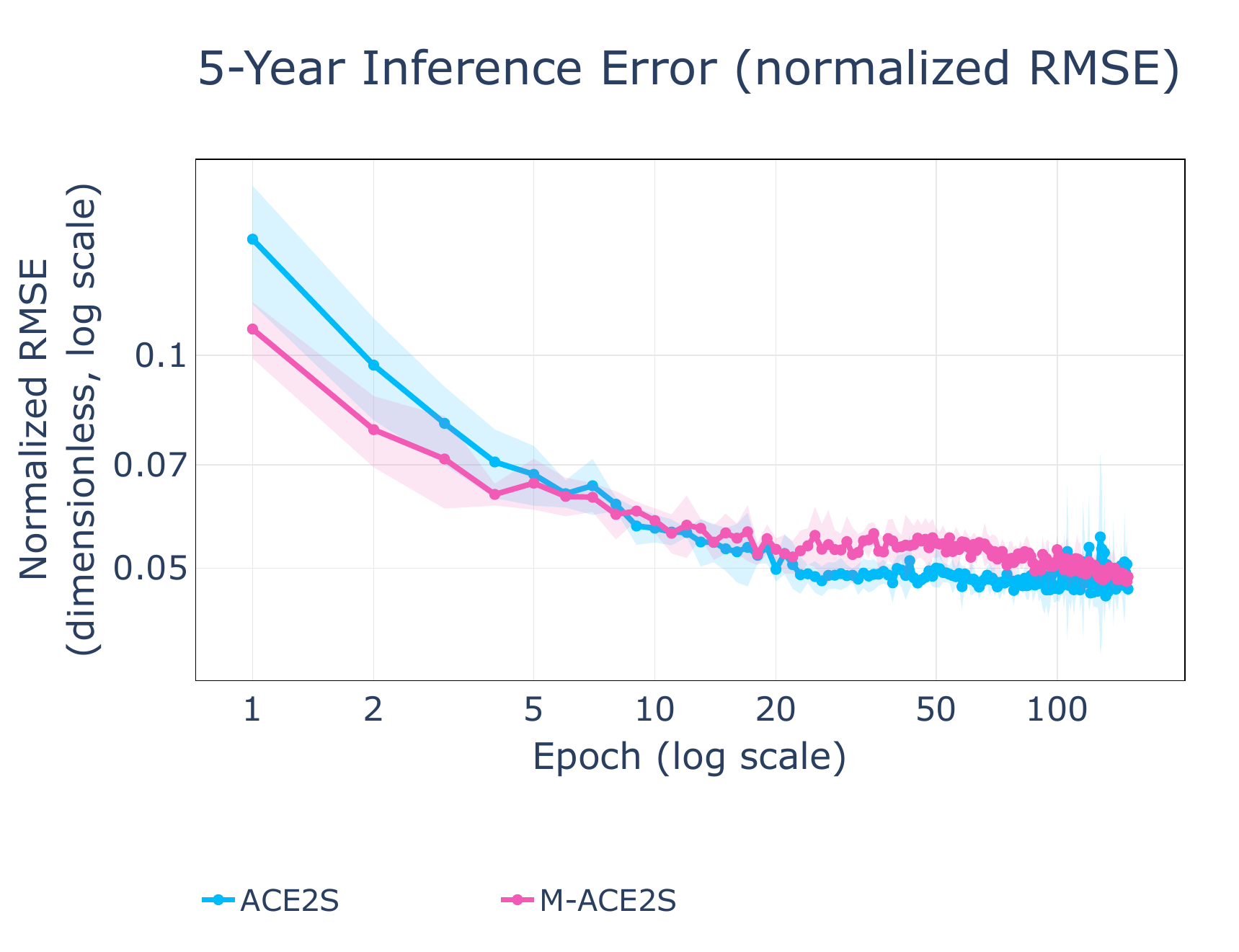}
  \caption{Training (dashed) and validation (solid) loss (left) and AIMIP checkpoint time-mean normalized RMSE averaged over channels (right) during training for M-ACE2S (pink) and ACE2S (blue) trained \emph{without} global mean removal (GMR off). Mean and standard deviation are computed over the ensemble of 5 random seed runs. Results are for 4\textdegree\ daily models; see \autoref{fig:46yr_era5_training_metrics_1deg_gmroff} for the 1\textdegree\ counterpart.}
  \label{fig:46yr_era5_training_metrics_4deg_gmroff}
\end{figure}

Here we provide the 4\textdegree\ counterparts of the training-metric curves shown for the 1\textdegree\ models in \autoref{fig:46yr_era5_training_metrics_1deg_gmron} and \autoref{fig:46yr_era5_training_metrics_1deg_gmroff}. Results are shown in \autoref{fig:46yr_era5_training_metrics_4deg_gmron} and \autoref{fig:46yr_era5_training_metrics_4deg_gmroff}. While the smaller gap between training and validation loss is preserved for 4\textdegree\ models, the reduction in final 5-year inference error is not.

\begin{thebibliography}{22}
\providecommand{\natexlab}[1]{#1}
\providecommand{\url}[1]{\texttt{#1}}
\expandafter\ifx\csname urlstyle\endcsname\relax
  \providecommand{\doi}[1]{doi: #1}\else
  \providecommand{\doi}{doi: \begingroup \urlstyle{rm}\Url}\fi

\bibitem[Henn et~al.(2026)Henn, Bretherton, Koldunov, Lessig, Molina, Arcomano,
  Watt-Meyer, Couairon, Singh, Brunstein, et~al.]{henn2026aimip}
Brian Henn, Christopher~S Bretherton, Nikolay Koldunov, Christian Lessig,
  Maria~J Molina, Troy Arcomano, Oliver Watt-Meyer, Guillaume Couairon, Renu
  Singh, Robert Brunstein, et~al.
\newblock {AIMIP} {Phase} 1: systematic evaluations of {AI} weather and climate
  models.
\newblock \emph{arXiv preprint arXiv:2605.06944}, 2026.

\bibitem[Gates et~al.(1999)Gates, Boyle, Covey, Dease, Doutriaux, Drach,
  Fiorino, Gleckler, Hnilo, Marlais, Phillips, Potter, Santer, Sperber, Taylor,
  and Williams]{gates1999amip}
W.~Lawrence Gates, James~S. Boyle, Curt Covey, Clyde~G. Dease, Charles~M.
  Doutriaux, Robert~S. Drach, Michael Fiorino, Peter~J. Gleckler, Justin~J.
  Hnilo, Susan~M. Marlais, Thomas~J. Phillips, Gerald~L. Potter, Benjamin~D.
  Santer, Kenneth~R. Sperber, Karl~E. Taylor, and Dean~N. Williams.
\newblock An overview of the results of the {Atmospheric Model Intercomparison
  Project} ({AMIP I}).
\newblock \emph{Bulletin of the American Meteorological Society}, 80\penalty0
  (1):\penalty0 29--56, 1999.
\newblock \doi{10.1175/1520-0477(1999)080<0029:AOOTRO>2.0.CO;2}.

\bibitem[Compo and Sardeshmukh(2009)]{compo2009oceanic}
Gilbert~P. Compo and Prashant~D. Sardeshmukh.
\newblock Oceanic influences on recent continental warming.
\newblock \emph{Climate Dynamics}, 32:\penalty0 333--342, 2009.
\newblock \doi{10.1007/s00382-008-0448-9}.

\bibitem[Dong et~al.(2009)Dong, Gregory, and Sutton]{dong2009understanding}
Buwen Dong, Jonathan~M. Gregory, and Rowan~T. Sutton.
\newblock Understanding land--sea warming contrast in response to increasing
  greenhouse gases. {Part I}: Transient adjustment.
\newblock \emph{Journal of Climate}, 22\penalty0 (11):\penalty0 3079--3097,
  2009.
\newblock \doi{10.1175/2009JCLI2652.1}.

\bibitem[Gregory and Andrews(2016)]{gregory2016variation}
Jonathan~M. Gregory and Timothy Andrews.
\newblock Variation in climate sensitivity and feedback parameters during the
  historical period.
\newblock \emph{Geophysical Research Letters}, 43\penalty0 (8):\penalty0
  3911--3920, 2016.
\newblock \doi{10.1002/2016GL068406}.

\bibitem[Andrews et~al.(2018)Andrews, Gregory, Paynter, Silvers, Zhou,
  Mauritsen, Webb, Armour, Forster, and Titchner]{andrews2018accounting}
Timothy Andrews, Jonathan~M. Gregory, David Paynter, Levi~G. Silvers, Chen
  Zhou, Thorsten Mauritsen, Mark~J. Webb, Kyle~C. Armour, Piers~M. Forster, and
  Holly Titchner.
\newblock Accounting for changing temperature patterns increases historical
  estimates of climate sensitivity.
\newblock \emph{Geophysical Research Letters}, 45\penalty0 (16):\penalty0
  8490--8499, 2018.
\newblock \doi{10.1029/2018GL078887}.

\bibitem[Watt-Meyer et~al.(2025)Watt-Meyer, Henn, McGibbon, Clark, Kwa,
  Perkins, Wu, Harris, and Bretherton]{watt2025ace2}
Oliver Watt-Meyer, Brian Henn, Jeremy McGibbon, Spencer~K Clark, Anna Kwa,
  W~Andre Perkins, Elynn Wu, Lucas Harris, and Christopher~S Bretherton.
\newblock {ACE2}: accurately learning subseasonal to decadal atmospheric
  variability and forced responses.
\newblock \emph{npj Climate and Atmospheric Science}, 8\penalty0 (1):\penalty0
  205, 2025.

\bibitem[Kochkov et~al.(2024)Kochkov, Yuval, Langmore, Norgaard, Smith, Mooers,
  Kl{\"o}wer, Lottes, Rasp, D{\"u}ben, et~al.]{kochkov2024neural}
Dmitrii Kochkov, Janni Yuval, Ian Langmore, Peter Norgaard, Jamie Smith,
  Griffin Mooers, Milan Kl{\"o}wer, James Lottes, Stephan Rasp, Peter
  D{\"u}ben, et~al.
\newblock Neural general circulation models for weather and climate.
\newblock \emph{Nature}, 632\penalty0 (8027):\penalty0 1060--1066, 2024.

\bibitem[Couairon et~al.(2024)Couairon, Singh, Charantonis, Lessig, and
  Monteleoni]{couairon2024archesweather}
Guillaume Couairon, Renu Singh, Anastase Charantonis, Christian Lessig, and
  Claire Monteleoni.
\newblock {ArchesWeather} \& {ArchesWeatherGen}: a deterministic and generative
  model for efficient {ML} weather forecasting.
\newblock \emph{arXiv preprint arXiv:2412.12971}, 2024.

\bibitem[Cresswell-Clay et~al.(2025)Cresswell-Clay, Liu, Durran, Liu, Espinosa,
  Moreno, and Karlbauer]{cresswell2025deep}
Nathaniel Cresswell-Clay, Bowen Liu, Dale~R Durran, Zihui Liu, Zachary~I
  Espinosa, Raul~A Moreno, and Matthias Karlbauer.
\newblock A deep learning {Earth} system model for efficient simulation of the
  observed climate.
\newblock \emph{AGU Advances}, 6\penalty0 (4):\penalty0 e2025AV001706, 2025.

\bibitem[Clark et~al.(2026)Clark, Arcomano, Duncan, Henn, Kwa, McGibbon,
  Perkins, Wu, Harris, Watt-Meyer, and Bretherton]{clark2026disentangling}
Spencer~K Clark, Troy Arcomano, James P~C Duncan, Brian Henn, Anna Kwa, Jeremy
  McGibbon, W~Andre Perkins, Elynn Wu, Lucas~M Harris, Oliver Watt-Meyer, and
  Christopher~S Bretherton.
\newblock Disentangling the effects of sea surface temperature and {CO$_2$} in
  global machine learned weather-climate emulators.
\newblock \emph{arXiv preprint arXiv:2606.07928}, 2026.

\bibitem[Clark et~al.(2025)Clark, Watt-Meyer, Kwa, McGibbon, Henn, Perkins, Wu,
  Harris, and Bretherton]{clark2025ace2}
Spencer~K Clark, Oliver Watt-Meyer, Anna Kwa, Jeremy McGibbon, Brian Henn,
  W~Andre Perkins, Elynn Wu, Lucas~M Harris, and Christopher~S Bretherton.
\newblock {ACE2-SOM}: Coupling an {ML} atmospheric emulator to a slab ocean and
  learning the sensitivity of climate to changed {CO$_2$}.
\newblock \emph{Journal of Geophysical Research: Machine Learning and
  Computation}, 2\penalty0 (4):\penalty0 e2024JH000575, 2025.

\bibitem[Brenowitz et~al.(2025)Brenowitz, Ge, Subramaniam, Manshausen, Gupta,
  Hall, Mardani, Vahdat, Kashinath, and Pritchard]{brenowitz2025climate}
Noah~D Brenowitz, Tao Ge, Akshay Subramaniam, Peter Manshausen, Aayush Gupta,
  David~M Hall, Morteza Mardani, Arash Vahdat, Karthik Kashinath, and Michael~S
  Pritchard.
\newblock {Climate in a bottle}: Towards a generative foundation model for the
  kilometer-scale global atmosphere.
\newblock \emph{arXiv preprint arXiv:2505.06474}, 2025.

\bibitem[Perkins et~al.(2025)Perkins, Kwa, McGibbon, Arcomano, Clark,
  Watt-Meyer, Bretherton, and Harris]{perkins2025hiro}
W~Andre Perkins, Anna Kwa, Jeremy McGibbon, Troy Arcomano, Spencer~K Clark,
  Oliver Watt-Meyer, Christopher~S Bretherton, and Lucas~M Harris.
\newblock {HiRO-ACE}: Fast and skillful {AI} emulation and downscaling trained
  on a 3 km global storm-resolving model.
\newblock \emph{arXiv preprint arXiv:2512.18224}, 2025.

\bibitem[Wu et~al.(2026)Wu, Duncan, Arcomano, McGibbon, Watt-Meyer, Bretherton,
  Mahfouz, Tebaldi, Van~Roekel, Roberts, et~al.]{wu2026stochastic}
Elynn Wu, James~PC Duncan, Troy Arcomano, Jeremy McGibbon, Oliver Watt-Meyer,
  Christopher~S Bretherton, Naser Mahfouz, Claudia Tebaldi, Luke Van~Roekel,
  Andrew Roberts, et~al.
\newblock Stochastic emulation of a fully coupled preindustrial {E3SMv3}
  simulation.
\newblock \emph{arXiv preprint arXiv:2608.10277}, 2026.

\bibitem[Hersbach et~al.(2020)Hersbach, Bell, Berrisford, Hirahara,
  Hor{\'{a}}nyi, Mu{\~{n}}oz-Sabater, Nicolas, Peubey, Radu, Schepers, Simmons,
  Soci, Abdalla, Abellan, Balsamo, Bechtold, Biavati, Bidlot, Bonavita, Chiara,
  Dahlgren, Dee, Diamantakis, Dragani, Flemming, Forbes, Fuentes, Geer,
  Haimberger, Healy, Hogan, H{\'{o}}lm, Janiskov{\'{a}}, Keeley, Laloyaux,
  Lopez, Lupu, Radnoti, Rosnay, Rozum, Vamborg, Villaume, and
  Th{\'{e}}paut]{Hersbach2020}
Hans Hersbach, Bill Bell, Paul Berrisford, Shoji Hirahara, Andr{\'{a}}s
  Hor{\'{a}}nyi, Joaqu{\'{\i}}n Mu{\~{n}}oz-Sabater, Julien Nicolas, Carole
  Peubey, Raluca Radu, Dinand Schepers, Adrian Simmons, Cornel Soci, Saleh
  Abdalla, Xavier Abellan, Gianpaolo Balsamo, Peter Bechtold, Gionata Biavati,
  Jean Bidlot, Massimo Bonavita, Giovanna Chiara, Per Dahlgren, Dick Dee,
  Michail Diamantakis, Rossana Dragani, Johannes Flemming, Richard Forbes,
  Manuel Fuentes, Alan Geer, Leo Haimberger, Sean Healy, Robin~J. Hogan,
  El{\'{\i}}as H{\'{o}}lm, Marta Janiskov{\'{a}}, Sarah Keeley, Patrick
  Laloyaux, Philippe Lopez, Cristina Lupu, Gabor Radnoti, Patricia Rosnay,
  Iryna Rozum, Freja Vamborg, Sebastien Villaume, and Jean-Noël Th{\'{e}}paut.
\newblock The {ERA}5 global reanalysis.
\newblock \emph{Quarterly Journal of the Royal Meteorological Society},
  146\penalty0 (730):\penalty0 1999--2049, June 2020.
\newblock \doi{10.1002/qj.3803}.

\bibitem[Bonev et~al.(2023)Bonev, Kurth, Hundt, Pathak, Baust, Kashinath, and
  Anandkumar]{bonev2023spherical}
Boris Bonev, Thorsten Kurth, Christian Hundt, Jaideep Pathak, Maximilian Baust,
  Karthik Kashinath, and Anima Anandkumar.
\newblock Spherical {Fourier} neural operators: Learning stable dynamics on the
  sphere.
\newblock In \emph{International Conference on Machine Learning}, pages
  2806--2823. PMLR, 2023.

\bibitem[Watt-Meyer et~al.(2023)Watt-Meyer, Dresdner, McGibbon, Clark, Henn,
  Duncan, Brenowitz, Kashinath, Pritchard, Bonev, et~al.]{watt2023ace}
Oliver Watt-Meyer, Gideon Dresdner, Jeremy McGibbon, Spencer~K Clark, Brian
  Henn, James Duncan, Noah~D Brenowitz, Karthik Kashinath, Michael~S Pritchard,
  Boris Bonev, et~al.
\newblock {ACE}: A fast, skillful learned global atmospheric model for climate
  prediction.
\newblock \emph{arXiv preprint arXiv:2310.02074}, 2023.

\bibitem[Liu et~al.(2025)Liu, Aksu, Liu, Liu, Yan, Pham, Savarese, Sahoo,
  Xiong, and Li]{liu2025moirai}
Chenghao Liu, Taha Aksu, Juncheng Liu, Xu~Liu, Hanshu Yan, Quang Pham, Silvio
  Savarese, Doyen Sahoo, Caiming Xiong, and Junnan Li.
\newblock {Moirai} 2.0: When less is more for time series forecasting.
\newblock \emph{arXiv preprint arXiv:2511.11698}, 2025.

\bibitem[Nie et~al.(2023)Nie, Nguyen, Sinthong, and Kalagnanam]{nie2022time}
Yuqi Nie, Nam~H Nguyen, Phanwadee Sinthong, and Jayant Kalagnanam.
\newblock A time series is worth 64 words: Long-term forecasting with
  transformers.
\newblock In \emph{International Conference on Learning Representations}, 2023.

\bibitem[{Lorenz}(1963)]{Lorenz1963deterministic}
Edward~N. {Lorenz}.
\newblock {Deterministic Nonperiodic Flow.}
\newblock \emph{Journal of the Atmospheric Sciences}, 20\penalty0 (2):\penalty0
  130--148, March 1963.
\newblock \doi{10.1175/1520-0469(1963)020<0130:DNF>2.0.CO;2}.

\bibitem[Slingo and Palmer(2011)]{slingo2011uncertainty}
Julia Slingo and Tim Palmer.
\newblock Uncertainty in weather and climate prediction.
\newblock \emph{Philosophical transactions. Series A, Mathematical, physical,
  and engineering sciences}, 369\penalty0 (1956):\penalty0 4751, 2011.

\end{thebibliography}
\end{document}